\documentstyle[11pt, epsf]{article}
\newcommand{\bra}[1]{\mbox{$\langle #1 |$}}
\newcommand{\ket}[1]{\mbox{$| #1 \rangle$}}
\newcommand{\braket}[2]{\mbox{$\langle #1  | #2 \rangle$}}
\newcommand{\bracket}[2]{\mbox{$\langle {{#1}} \mathrel{ | {\vphantom
        {{#1} {#2}}} \kern-\nulldelimiterspace} {{#2}} \rangle$}}
\newcommand{\proj}[1]{\ket{#1}\!\bra{#1}}
\newcommand{\rem}[1]{}

\newcommand{\ros}{\rho_s}
\newcommand{\rost}{\rho_s^{T_B}}
\newcommand{\rmes}{\rho_s^{+}}

\newcommand{\rmenys}{\rho_s^{-}}

\def\un{\leavevmode\hbox{\normalsize1\kern-4.6pt\large1}}
\def\un{\hbox{$\mit I$\kern-.277em$\mit I$}}

\begin{document}

\title {Robustness of entanglement}
\author{Guifr\'e Vidal and Rolf Tarrach
\\
Departament d'Estructura i Constituents de la Materia, \\
Universitat de Barcelona, 08028 Barcelona, Spain.}

\maketitle

\begin{abstract}
 In the quest of completely describing entanglement in the general case of a finite number of parties sharing a physical system of finite dimensional Hilbert space a new entanglement magnitude is introduced for its pure and mixed states: robustness. It corresponds to the minimal amount of mixing with locally prepared states which washes out all entanglement. It quantifies in a sense the endurence of entanglement against noise and jamming. Its properties are studied comprehensively. Analytical expressions for the robustness are given for pure states of binary systems, and analytical bounds for mixed states of binary systems. Specific results are obtained mainly for the qubit-qubit system. As byproducts local pseudomixtures are generalized, a lower bound for the relative volume of separable states is deduced and arguments for considering convexity a necessary condition of any entanglement magnitude are put forward.
\end{abstract}

\vspace{4mm}
{\bf 1. INTRODUCTION.}
\vspace{4mm}

 Entanglement \cite{Schro}, \cite{EPR} is arguably the most non-classical feature of quantum mechanics. For it to show up, the physical system has to consist of different local parts, which we will call {\em local subsystems}, in one-to-one correspondence with different physicists, which we will call {\em parties}. Each party acts locally on its respective local subsystem. Each local subsystem will, in general, consist of further parts, {\em local partial subsystems} or {\em objects}, which may be entangled among themselves, locally. We are only concerned here with non-local entanglement, involving more than one local subsystem. The system might also be partitioned into {\em non-local subsystems}, which are shared by several parties (see Appendix A).

 Entanglement refers thus to states shared by more than one party. It is behind (or depending on the definitions, equivalent to) non-locality, non-separability and the existence of non-classical or quantum correlations, as seen by the parties. It plays a central role in quantum communication and quantum computation. A huge effort is being put into quantifying entanglement. This is an extremely difficult undertaking, mainly because of the intricate interplay between classical and quantum correlations. What one would like to have is a minimal set of independent, physically meaningful magnitudes, which completely characterize entanglement.

 Consider e.g. a binary system consisting of two three-state local subsystems (each one being, say, a spin 1 particle). As far as entanglement is concerned, any pure state of this system is completely determined by two of the three coefficients of its Schmidt decomposition \cite{Schmidt}:

\begin{equation}
\ket{\Psi} = a_1\ket{1}\!\otimes\!\ket{1} + a_2\ket{2}\!\otimes\!\ket{2} + a_3\ket{3}\!\otimes\!\ket{3}, \,\,\,\,\, a_i \geq 0, \,\,\, \sum_{i=1}^3 a_i^2 =1,
\end{equation}
say, the two largest ones. Thus two independent magnitudes will suffice to completely characterize its entanglement. Suppose one chooses one of them to be the entropy of entanglement \cite{BBPSch},

\begin{equation}
E(\Psi) = - \sum_{i=1}^3 a_i^2\log_2 a_i^2.
\label{entropy}
\end{equation}
 Consider now two pure states with different Schmidt coefficients, $\ket{\Psi_1} \neq \ket{\Psi_2}$, which have the same entropy of entanglement, $E(\Psi_1) = E(\Psi_2)$. As we will see, there are physically meaningful entanglement magnitudes which quantify differently these two states. Any such magnitude, together with the entropy of entanglement we chose to start with, determines the two largest Schmidt coefficients and thus characterizes entanglement for pure states completely (provided they are bijective). 

 As the local subsystems become more complex involving more states and more partial subsystems, as the number of subsystems and thus parties grows, so does the number of entanglement magnitudes needed to completely characterize entanglement \cite{LinPop}.

 The measures of entanglement proposed up to now are examples of entanglement magnitudes. Foremost is the entanglement of formation or creation \cite{BdVSW}. Others are the entanglement of distillation  \cite{BdVSW} and the relative entropy of entanglement  \cite{VedralPlenio}, but several more have been proposed recently (e.g. entanglement of assistance  \cite{dVFMSTU}). It has been argued that for pure states there is a unique measure of entanglement  \cite{PopRoh}, but certainly one sole magnitude will, in general, not be enough for characterizing entanglement completely.

 The aim of this contribution is to propose a new entanglement magnitude which we will call {\em robustness} and to study it in some detail. It has several appealing features. Its definition is simple and valid for any state of a composite system composed by any finite number of local subsystems of finite dimension. It is based on a simple physical operation: mixing with locally prepared states. It does not increase on average when the parties, classically communicated, act locally on the subsystems. The {\em robustness} quantifies the endurance of entanglement with respect to local mixing by asking about the minimal amount of entanglement-free mixing needed to wash out all entanglement. It can be interpreted as a quantification of intelligent jamming of entanglement, intelligent meaning that the parties know the entangled state and thus tailor the jamming accordingly so that a minimal amount suffices. An auxiliary and useful magnitude will be the {\em random robustness}, which can be interpreted as the robustness of entanglement with respect to mixing with white noise. While we explain and analyze robustness a few more general results will be presented: convexity of entanglement magnitudes is put on firmer grounds,  a set of necessary and sufficient conditions for consistency with the fundamental law of quantum information processing is presented \cite{VP2} and a weak version of the composition law of entanglement magnitudes for a system which consists of two uncorrelated non-local entangled subsystems is suggested.

 The paper is organized as follows. In section 2, after analyzing some features of mixing, we introduce, following  \cite{STV}, {\em local pseudomixtures} and prove their existence for the most general, finite dimensional, case, thus generalizing local descriptions of entanglement. A universal local pseudomixture is given for any state of this general case. {\em Relative robustness} and {\em random robustness} are also introduced and their physical meaning discussed. In section 3 we introduce {\em robustness}, and prove eight general properties that make it a potentially useful entanglement magnitude. In section 4 a number of results for the robustness of binary systems are presented, whose proofs can be found in Appendices B and C. They include explicit expressions for the robustness and the random robustness of any pure state, and bounds for mixed states. For the two simplest binary systems more accurate results are presented and a numerical method for the computation of the robustness is discussed. We present an application of some of the results of the previous section in section 5 by obtaining a universal lower bound of the relative volume of separable states, completing thus ref. \cite{volume}. 

 One of the main questions concerning entanglement measures is, what are the necessary and what are the sufficient conditions they have to fulfill? Lot of progress has been achieved in the last years (see e.g. \cite{VedralPlenio}, \cite{PopRoh})  although many questions still remain, in particular concerning additivity. We hope this and further studies of robustness will also contribute to the understanding of these issues.

\vspace{4mm}

{\bf 2. LOCAL PSEUDOMIXTURES.}

\vspace{4mm}

{\bf 2.A Mixing of shared states and local operations.}
\vspace{4mm}

 Since the mixing of states that are shared by several parties will play a major role throughout this contribution, we find it convenient to begin with a few comments on how one can obtain a density matrix from one of its {\em realizations}, without resorting to non-local operations. This will as a byproduct lead to a new condition any measure of entanglement has to satisfy.

 We will call a set of states $\{\rho_k\}_{k=1 \cdots l}$ with associated probabilities $\{p_k\}_{k=1\cdots l}$ a realization \footnote{
 We will not call it ensemble because we construe from it the state $\rho$ by dismissing information, not by choosing randomly one item out of an ensemble of states $\rho_k$ populated proportionally to $p_k$.
}
 $\Upsilon \equiv \{\rho_k,p_k\}_{k=1\cdots l}$ of the density matrix $\rho \equiv \sum_{k=1}^l p_k\rho_k$. A device $\Sigma$ that is known to supply a system ${\cal Q}$ in the state $\rho_k$ with probability $p_k$, for $k=1\cdots l$, provides the parties with the system ${\cal Q}$ in a state that depends on the amount of extra information $\Sigma$ supplies together with the prepared system ${\cal Q}$. Thus, if $a)$ no extra information is supplied, then the state of ${\cal Q}$ is $\rho$, whereas this is not the case if $b)$ $\Sigma$ casts a message stating which specific state $\rho_k$ ${\cal Q}$ has been prepared in, such an event having an $a$ $priori$ probability $p_k$. These two situations lead to states of ${\cal Q}$ that are clearly inequivalent even in a statistical sense, and this fact is exemplified if one adopts an utilitarian approach: consider any function $\mu(\rho)$ defined on the set of states that quantifies somehow some resources contained in $\rho$. (Alternatively $\mu(\rho)$ could quantify the cost of preparing the state $\rho$, and so on). Then, in situation $a)$ the parties obtain a state $\rho$, from which they can extract, maybe after some manipulations, an amount $\mu(\rho)$ of resources, whereas in situation $b)$ the expected amount of resources the parties can extract is an average, over the realization $\Upsilon$, of the amounts $\mu(\rho_k)$, i.e. $\mu(\Upsilon) \equiv \sum_{k=1}^l p_k\mu(\rho_k)$, the extra information supplied together with ${\cal Q}$ allowing for a conditional treatment of this system depending on the concrete $\rho_k$ the parties get. Moreover, whatever is done to ${\cal Q}$ in situation $a)$ in order to use it as a resource, the very same manipulations can be done in $b)$ regardless of the extra information supplied, obtaining, in a statistical sense, the same results as in $a)$, so that one gets, {\em on average}, at least as many resources in case $b)$ as in case $a)$. Therefore 
\begin{equation}
\mu(\rho) \leq \mu(\Upsilon). 
\end{equation} 
(Were $\mu(\rho)$ a quantification of the minimal cost of preparation of $\rho$, one could reach the same conclusion by noticing that $\Upsilon$ is not the only realization which leads to $\rho$, and that there may be cheaper ones, that is $\mu(\Upsilon) \geq \min_{\Upsilon'} \mu(\Upsilon')\,\, (\equiv \mu(\rho))$, where $\Upsilon'$ is any realization of $\rho$.) 
 
 Notice, moreover, that since the only difference between situation $a)$ and $b)$ consists on the extra information supplied in $b)$, if this extra information is irreversibly lost for the parties the largest amount of resources that they can obtain from ${\cal Q}$ becomes $\mu(\rho)$, even if initially the expected amount has been $\mu(\Upsilon)$.

 Let us translate the above considerations to the case where $\mu$ is any measure of entanglement, which we will do with a concrete example. Suppose $\Sigma$ prepares two particles in the global state $\rho_k$ with probability $p_k$ and then sends one to Alice and the other to Bob (and thus ${\cal Q}$ is a binary system). What we want to remark here is that the loss of the extra information supplied in case $b)$, which forces a transition of the state of ${\cal Q}$ from a $\rho_k$ to $\rho$, can occur without Alice and Bob having to put the two particles back together, so that to all effects it can be regarded as a local process. Then, in particular, we have argued that any measure  ${\cal E}$ of the entanglement of a shared state $\rho$ has to be a convex function (see also \cite{VedralPlenio}), that is
\begin{equation}
{\cal E}(\rho) \leq \sum_{k=1}^l p_k{\cal E}(\rho_k) 
\end{equation} 
if $\rho = \sum_{k=1}^l p_k\rho_k$, otherwise one would be creating, on average, entanglement by means of local operations (as loosing information is something one can always do locally). Notice that the entanglement of assistance \cite{dVFMSTU} is concave, not convex (it also is generally non-vanishing for separable states). We interpret it as a measure for ensembles of maximally entangled pure states realizing a density matrix, not as a proper measure for a single system described by the density matrix.

 Once a possible way of (locally) obtaining a state from any of its information supplied realizations has been discussed, we would like to address a question motivated by the fact that a mixture of shared states, even if they all are entangled, may contain no entanglement at all, so that the procedure of mixing often implies the disappearance of quantum correlations: specifically, given an arbitrary entangled state of a composite system shared by $N$ parties, we would like to know whether it is always possible to wash out all its quantum correlations by mixing it with an adequate separable state. This will be our starting point to derive an entanglement magnitude: the {\em robustness} of entangled states.

\vspace{3mm} 
{\bf 2.B Erasing quantum correlations by mixing with a separable state: a local description of entangled states.}
\vspace{2mm} 

 Consider a composite system ${\cal Q}$ with $N$ local subsystems such that the dimension of its Hilbert space ${\cal H}$, $n$, is finite. Let us recall that, according to whether they can be expressed as a convex combination of pure product states or not, one can distinguish between separable and entangled states. Thus, for $\{{\cal H}^i\}_{i=1,...N}$ the Hilbert spaces of the local subsystems (${\cal H} = \bigotimes_{i=1}^N {\cal H}^i$), separable states $\ros$ can be written as
\begin{equation}
\ros = \sum_{k} p_k \proj{\Psi_k},
\label{separable}
\end{equation} 
where $p_k > 0, \;\: \sum_{k} p_k = 1$ and $\ket{\Psi_k} = \bigotimes_{i=1}^N \ket{\Psi^i_k} \in {\cal H}, \;\; \ket{\Psi^i_k} \in {\cal H}^i$.
We will now introduce the concept of robustness of a state $\rho \in {\cal T}({\cal H})$ relative to a separable state $\ros \in {\cal S}({\cal H})$ (by ${\cal T}({\cal H})$ we denote the set of states of ${\cal Q}$, and by ${\cal S}({\cal H})\subset {\cal T}({\cal H})$ that of separable states of the same system).

\vspace{3mm} 
{\bf Definition:} Given a state $\rho \in {\cal T}({\cal H})$ and a separable state $\ros \in {\cal S}({\cal H})$, we call {\em robustness} of $\rho$ {\em relative} to $\ros$, $R(\rho||\ros)$, the minimal $s \geq 0$ for which 
\begin{equation}
\rho(s) \equiv \frac{1}{1+s}(\rho + s\ros)
\label{relative}
\end{equation}
is separable. 

We will single out a particular case of the relative robustness by giving it its own name. 

\vspace{3mm} 
{\bf Definition:} We call {\em random robustness} of $\rho$ its robustness relative to the (separable) maximally random state $\frac{1}{n}I$.

Thus $R(\rho||\ros)$ is the minimal amount of $\ros$ that has to be mixed with $\rho$ in order to wash out all the quantum correlations initially contained in $\rho$. Notice that $R(\rho||\ros)$ is zero if, and only if, $\rho$ is separable itself. Our previous question, which Theorem 1 will answer, reduces now to see whether one can always find a separable $\ros$ such that $\rho$ has finite relative robustness $R(\rho||\ros)$. Equivalently, in terms of the local pseudomixtures introduced in  \cite{STV}, we would like to know whether one can always express a state $\rho \in {\cal T}({\cal H})$ as 
\begin{equation}
\rho = (1+t)\rmes - t\rmenys, \;\;\;\; \; 0 \leq t < \infty,
\label{pseudo}
\end{equation}
for some $\rmes,\,\rmenys \in {\cal S}({\cal H})$, that is, whether one can always describe a state as a local pseudomixture (see \cite{Kuna} for a recent proof for $N=2$). Notice that expressing $\rmes$ and $\rmenys$ in Eq.(\ref{pseudo}) as {\em finite} statistical mixtures of pure product states $\ket{\Psi_k}=\bigotimes_{i=1}^N \ket{\Psi}, \;\; \ket{\Psi^i_k} \in {\cal H}^i$  \cite{Horo}, one gets
\begin{equation}
\rho = \sum_{k=1}^{l < \infty} r_k \proj{\Psi_k} 
\label{pseudo2}
\end{equation}
where $\sum_{k=1}^{l < \infty} r_k = 1$ and $r_k \in {\cal R}$. That is, a state $\rho$ of ${\cal Q}$ is expressed as a sum of pure product states, in a similar way to how separable states are as statistical mixtures, but with the difference that now the probabilistic weights $p_k$ in Eq.(\ref{separable}) have been replaced by real numbers $r_k$, still restricted by $\sum_{k=1}^{l < \infty} r_k = 1$. It is this resemblance with mixtures what motivates calling the right hand sides of Eqs. (\ref{pseudo}) and (\ref{pseudo2}) {\em local pseudomixtures}, the adjective {\em local} reflecting the fact that all states intervening in such expressions are separable.

\begin{figure}
 \epsfysize=6.0cm
 \epsffile{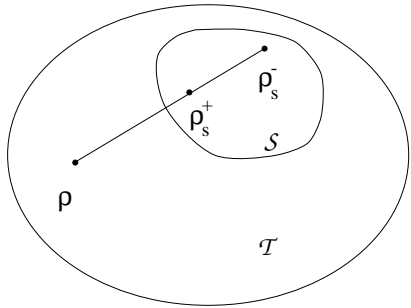}
 \caption{Local pseudomixture for the entangled state $\rho$. Since there always exists a $\rmenys \in {\cal S}$ and a finite $t>0$ such that $\rmes\equiv\frac{1}{1+t}(\rho + t\rmenys)$ belongs to ${\cal S}$, one can express $\rho$ in terms of two separable states and the weight $t$ as $\rho = (1+t)\rmes - t\rmenys$.}
\label{figura1}
\end{figure}

\begin{figure}
 \epsfysize=6.0cm
 \epsffile{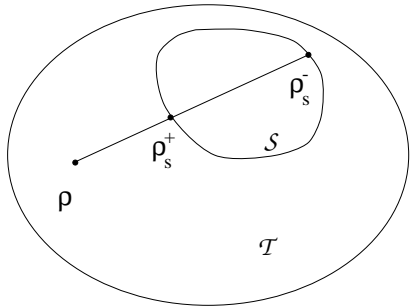}
 \caption{An optimal local pseudomixture for the state $\rho$ is such that the weight $t=R(\rho||\rmenys)$ is minimal. Thus the robustness $R(\rho||{\cal S})$ is a geometrical quantity that relates the element $\rho \in {\cal T}$ with the subset ${\cal S}\subset{\cal T}$, as $p = \frac{1}{1+R(\rho||{\cal S})}$ is the maximal weight of $\rho$ in a convex combination $p\rho + (1\!-\!p)\rmenys$ involving an element $\rmenys \in {\cal S}$ such that it belongs to the subset ${\cal S}$. }
\label{figura2}
\end{figure}

\vspace{3mm} 
{\bf Theorem 1}
\vspace{3mm} 

 Any entangled state $\rho$ of a generic composite system (with finite-dimensional Hilbert space) can be expressed in terms of two separable states and a non-negative finite real number, $\{\rmes,\rmenys, t\}$, as $\rho = (1+t)\rmes - t\rmenys$ (see Figure \ref{figura1}).
\vspace{3mm} 

{\bf Proof:} In Appendix C an explicit upper bound for the random robustness of any state $\rho$,
\begin{equation}
R(\rho||\frac{1}{n}I) \,\, \leq \,\, (1+\frac{n}{2})^{N\!-\!1} - 1 \equiv \tilde{t},
\end{equation}
will be obtained. Then one can write
\begin{equation}
\rho = (1+\tilde{t})\rmes - \tilde{t}\frac{1}{n}I, \,\,\,\,\,\,\, \rmes \equiv \frac{1}{1+\tilde{t}}(\rho + \tilde{t}\frac{1}{n}I),
\end{equation}
where $\rmenys$ in Eq.(\ref{pseudo}) is $\frac{1}{n}I$ and $\rmes$ is separable by construction. $\,\,\,\Box$

 Notice that this means, in particular, that $\frac{1}{n}I \in {\cal S}$ is not on the frontier of ${\cal S}$ and  ${\cal T}\backslash {\cal S}$, but in the interior of ${\cal S}$, as was first proved in \cite{volume}. Our proof, which is independent of that presented in  \cite{volume}, provides offhand an explicit pseudomixture for any state $\rho$ and implies, from a physical point of view, that one can always erase all quantum correlations by mixing with the maximally random state. Let us remark that, mixing with $\frac{1}{n}I$ can be regarded as a toy model for the study of the effect of white noise on the quantum correlations contained in an entangled state. We will come back to the study of the random robustness in section 4.B, where it will be computed for any pure state of a binary (i.e. $N=2$) system and for any state of binary systems of dimension $n\leq 6$, whereas for any mixed state of a generic binary system lower and upper bounds will be presented. Finally, the random robustness will be used in section 5 to obtain an explicit lower bound for the volume of separable states.   

\vspace{4mm} 
{\bf 3. ROBUSTNESS OF SHARED STATES.}
\vspace{4mm}

{\bf 3.A Definition of robustness.}
\vspace{4mm}

We have established so far that for any state $\rho$ there is at least one separable state $\rho_s$ such that $R(\rho||\ros)$ is finite. In addition we have seen that this $\ros$ can be chosen to be independent of $\rho$. We now prove the existence of a minimal value of $R(\rho||\ros)$ as a function of $\ros \in {\cal S}$. This quantity, $R(\rho||{\cal S})$, will prove to be, on average, non-increasing under any transformation of the shared system involving only local operations on the subsystems and classical communication between the parties. Analogously to $R(\rho||\ros)$, $R(\rho||{\cal S})$ is the minimal amount of any separable state that has to be mixed with $\rho$ in order to wash out its quantum correlations, and has a neat geometrical meaning (see Figure \ref{figura2}). Notice that no metric in ${\cal T}$ has been used to define $R(\rho||{\cal S})$.

 Let us consider, then, a state $\rho$ of ${\cal Q}$ and all its possible pseudomixtures \{$ \rmes,\, \rmenys,\, t\}$ of separable states.

\vspace{2mm}
{\bf Lemma 1}

 There always exists a non-negative $t(\rho; \rmes, \rmenys)$ satisfying Eq.(\ref{pseudo}) which is the minimal one.

{\bf Proof:} It follows from the fact that $(\rmes$, $\rmenys)$ must belong to a compact subset of ${\cal S}${\sf x}${\cal S}$ (since they are constrained by Eq.(\ref{pseudo})) and that $t(\rho; \rmes, \rmenys) \geq 0$ is a continuous function of them.$\,\,\,\Box$
\vspace{2mm}

{\bf Definition:} We call ({\em absolute}) {\em robustness} of $\rho \in {\cal T}$ the quantity
\begin{equation}
R(\rho||{\cal S}) \equiv \min_{\ros \in {\cal S}} R(\rho||\ros).
\end{equation}

\vspace{2mm}
{\bf Definition:} We call a local pseudomixture with $t(\rho; \rmes, \rmenys) = R(\rho||{\cal S})$ an {\em optimal} one. 
\vspace{2mm}

\vspace{3mm} 
{\bf 3.B Some properties of robustness.}
\vspace{3mm} 

 Next we will discuss eight properties the robustness of a state satisfies. Some of them are necessary to guarantee that an entanglement magnitude cannot be increased locally (that is, by means of the combined use of local transformations and classical communication) (cf. \cite{VedralPlenio}, \cite{PopRoh}). Another assures that the magnitude allows to distinguish between separable and entangled states. The last property is a weak version of a composition law, which replaces additivity.

 Recall that the robustness has been defined for states of a generic composite system, so that it can be applied to states shared by an unrestricted (but finite) number of parties $N$. We associate parties with local subsystems, so that a local subsystem consists of all the physical objects (particles, for instance) a party holds and can act on. We also require each of these local sets of objects to have a Hilbert space ${\cal H}^i$ of finite dimension $n_i$, so that $\rho$ will be a $n$x$n$ matrix acting on ${\cal H} = \bigotimes_{i=1}^N {\cal C}^{n_i}$, with $n = \dim({\cal H}) = \prod_{i=1}^N n_i$. In analogy with pure product states, product operators will be those that can be expressed as $O = \bigotimes_{i=1}^N O^i$, with $O^i$ an operator in ${\cal H}^i$, and a product subspace of ${\cal H}$ will be a space $\tilde{\cal H} \subseteq {\cal H}$ such that $\tilde{\cal H} = \bigotimes_{i=1}^N\tilde{\cal H}^i$ with $\tilde{\cal H}^i$ a subspace of ${\cal H}^i$. The robustness of a state, from now on simply $R(\rho)$, satisfies:
\vspace{3mm}

($i$) If range$(\rho) \subseteq \tilde{\cal H} \subset {\cal H}$, then $R(\rho)$ is independent of the Hilbert space $\tilde{\cal H}$ or ${\cal H}$ $\rho$ acts on.\footnote{The support or range of a density matrix is the subspace spanned by its eigenvectors of non-vanishing eigenvalue. The dimension of the range is the rank.}

($ii$) $R(\rho) \geq 0$; $R(\rho) = 0 \,\Leftrightarrow \, \rho \in {\cal S}$. 

($iii$) $R(\rho) = R(U_L \rho U_L^{\dagger})$ for any unitary product operator $U_L = \bigotimes_{i=1}^N U^i$.

($iv$) $R(Tr_{\tilde{\cal Q}}[\rho]) \leq  R(\rho)$, where $Tr_{\tilde{\cal Q}}[.]$ is a partial trace over $\tilde{\cal Q}$, $\tilde{\cal Q}$ denoting any subset, local or not, of the whole set of objects held by the parties.

($v$) $R(\rho\otimes\ros) = R(\rho)$, where $\ros$ is any separable state. 

($vi$) $R(\rho) \leq \sum_{k} p_k R(\rho_k)$, where $\{\rho_k, p_k\}$ is any realization of $\rho$, i.e. $\rho = \sum_{k} p_k \rho_k$. 

($vii$) $R(\rho) \geq \sum_{k} p_k R(\rho_k)$ if as a result of an incomplete local von Neumann measurement $\rho$ becomes the state $\rho_k$ with probability $p_k$. 

($viii$) $m(R(\rho_x),R(\rho_y))\,\leq \, R(\rho_x\otimes\rho_y) \, \leq \, M(R(\rho_x),R(\rho_y))$, where $m$ and $M$ are two functions of $R(\rho_x)$ and $R(\rho_y)$.

\vspace{3mm}

 The meaning of property ($i$) is that the robustness of a state is not an intensive quantity in the dimension $n$ of the Hilbert space of the shared system ${\cal Q}$ - since it is independent of $n$ -. The following example should clarify the meaning of ($i$): the two-party pure entangled state $\ket{\Psi} = \frac{1}{\sqrt{2}} (\ket{1}\!\otimes\!\ket{1} + \ket{2}\!\otimes\!\ket{2})$, where $\ket{1}$ and $\ket{2}$ are two normalised orthogonal vectors, has a density matrix $\proj{\Psi}$ that can act, for instance, on ${\cal C}^{2} \otimes {\cal C}^{2}$ (a two qubit system) or on ${\cal C}^{3} \otimes {\cal C}^{3}$ (a two qutrit system). What property ($i$) assures is that $R(\Psi)$ does not depend on whether $\ket{\Psi}$ is the state of two qubits or of two qutrits, and it is not obviously satisfied (later on, for instance, it will be seen that the random robustness of $\ket{\Psi}$, $R(\Psi||\frac{1}{n}I)$, does depend on $n$).

A generic $\rho$ in ${\cal H} = \bigotimes_{i=1}^N {\cal H}^i$ may have support only on a product subspace of ${\cal H}$. Let us call $\tilde{\cal H} \subseteq {\cal H}$ the smallest of such product subspaces ($\tilde{\cal H} = \bigotimes_{i=1}^N\tilde{\cal H}^i$ can be constructed by computing, for each party $i$, its local state $\rho^{i} =  Tr_{{\cal Q}\backslash{\cal Q}^i}[\rho]$, and by taking $\tilde{\cal H}^i \subseteq {\cal H}^i$ to be the subspace spanned by the eigenvectors of $\rho^{i}$ with non-vanishing eigenvalue). Taking up the previous example involving $\ket{\Psi} \in {\cal C}^{3} \otimes {\cal C}^{3}$, one can see that the projection of $\proj{\Psi}$ onto the product subspace $\tilde{\cal H} = < \ket{1},  \ket{2} > \otimes < \ket{1},  \ket{2} > \cong {\cal C}^{2} \otimes {\cal C}^{2}$ by means of the product projector $P = (\proj{1} + \proj{2})\otimes (\proj{1} + \proj{2})$ leaves the state unchanged, that is $P\proj{\Psi}P = \proj{\Psi}$, and no other product projector of equal or smaller rank does so.

 If we now show that any optimal local pseudomixture of $\rho$, $\{\rmes,\rmenys, t = R(\rho)\}$, satisfies that both $\rmes$ and $\rmenys$ have support only on $\tilde{\cal H}$, then it will be irrelevant, in terms of $R(\rho)$, whether $\rho$ acts on the whole ${\cal H}$ or only on $\tilde{\cal H}$.

\vspace{3mm} 
{\bf Theorem 2}
\vspace{3mm} 

 For any optimal pseudomixture of $\rho$, $\{\rmes,\rmenys, R(\rho)\}$, if $\tilde{\cal H}$ is the smallest product subspace supporting $\rho$ and $P$ is a projector onto it, then $P\rmes P = \rmes$ and $P\rmenys P = \rmenys$.

{\bf Proof:} Notice first that for a normalised pure product state $\ket{\Psi} = \bigotimes_{i=1}^N \ket{\Psi^i}$, its projection onto a product subspace is, if not zero, another pure product state $\ket{\Phi} = P\ket{\Psi} = \bigotimes_{i=1}^N P^i\ket{\Psi^i}=\bigotimes_{i=1}^N \ket{\Phi^i}$, with $\langle \Phi|\Phi \rangle \leq 1$. Then, since any separable state $\ros$ can be expressed as a convex combination of projectors onto pure product states $\ket{\Psi_k}$, i.e. $\ros = \sum_{k} p_k \proj{\Psi_k}$, and $P\proj{\Psi_k}P = \proj{\Phi_k} = q_k\proj{\tilde{\Phi}_k}$, where $0 < q_k \equiv \braket{\Phi_k}{\Phi_k} \leq 1$ and $\ket{\tilde{\Phi}_k} \equiv \frac{1}{\sqrt{q_k}} \ket{\Phi_k}$ is a normalised pure product state (unless $P\ket {\Psi_k} = 0$), the renormalised restriction of $\ros$ on 
$\tilde{\cal H}$, $\tilde{\rho}_s \equiv \frac{P\ros P}{Tr[P\ros P]}$, ($Tr[P\ros P] = \sum_k p_k q_k \leq 1$), is a separable density matrix as well. Then
\begin{equation}
\rho = P\rho P = (1+R(\rho)) Tr[P\rmes P] \tilde{\rho}_s^+ - R(\rho) Tr[P\rmenys P] \tilde{\rho}_s^-.
\end{equation}
 
 Now suppose that at least one of $\rmes$ and $\rmenys$ (and thus, in fact, both), say $\rmenys$, has support not contained in $\tilde{\cal H}$. Then $Tr[P\rmenys P] < 1$ and we automatically obtain a new local pseudomixture involving $\tilde{\rho}_s^- \neq \rmenys$, with $t = Tr[P\rmenys P]\, R(\rho)$$ < R(\rho)$, which is a contradiction, for we started from an optimal one.$\,\,\,\Box$
 
 Property ($ii$) says that the robustness of a state $\rho$ indicates whether $\rho$ is entangled or separable. To see ($ii$), notice that $R(\rho) = 0$ implies that $\rho = \rmes$, which is separable, and that if $\rho$ is separable, then by choosing $\rmes$ to be $\rho$ one gets a local pseudomixture for $\rho$, with $t = 0$.

Property ($iii$) states that any two states related by a unitary product transformation have the same robustness.

\vspace{2mm}
{\bf Theorem 3}
\vspace{2mm}

$R(\rho) = R(U_L \rho U_L^{\dagger})$.

{\bf Proof:}
Notice that $R(U_L \rho U_L^{\dagger})$ cannot be greater than $R(\rho)$, since by transforming an optimal local pseudomixture,
\begin{equation}
\rho = (1+R(\rho))\rmes - R(\rho)\rmenys,
\end{equation}
 by $U_L$ we find the local pseudomixture
\begin{equation}
U_L\rho U_L^{\dagger} = (1+R(\rho))U_L\rmes U_L^{\dagger} - R(\rho)U_L\rmenys U_L^{\dagger},
\end{equation}
which has $t=R(\rho)$. {\em Mutatis mutandis} we see that $R(U_L \rho U_L^{\dagger})$ cannot be smaller than $R(\rho)$.$\,\,\,\Box$

\vspace{2mm}

 In order to discuss properties  $(iv)$ and $(v)$, recall that if to ${\cal Q}$, in the state $\rho$, we add a set of objects $\tilde{\cal Q}$ in the state $\tilde{\rho}$, then the state of ${\cal Q}\bigcup\tilde{\cal Q}$ is $\rho\otimes\tilde{\rho}$ (assuming ${\cal Q}$ and $\tilde{\cal Q}$ are uncorrelated). On the other hand, for $\tilde{\cal Q} \subset {\cal Q}$ a subset of objects, if $\rho$ is the state of ${\cal Q}$ then that of $\tilde{\cal Q}$ is $Tr_{{\cal Q}\backslash\tilde{\cal Q}}[\rho]$, whereas if we throw $\tilde{\cal Q}$ away the remaining state is $Tr_{\tilde{\cal Q}}[\rho]$.

 Point $(iv)$ states that the robustness of the state $\rho$ of a composite system ${\cal Q}$ does not increase when throwing away any subset of objects $\tilde{\cal Q} \subset {\cal Q}$.

\vspace{2mm}
{\bf Theorem 4}
\vspace{2mm}

$R(Tr_{\tilde{\cal Q}}[\rho]) \leq R(\rho).$

{\bf Proof:} It will suffice to analise the case of $\tilde{\cal Q}$ being a single object held by one party, since for a general $\tilde{\cal Q} = \bigcup_{i, j} {\cal Q}^{i,j}$ one can proceed stepwise, each step involving only one object. Take then $\tilde{\cal Q} ={\cal Q}^{1,1}$ (relabelling the parties and objects, if necessary), so that the partial trace is taken over the factor space ${\cal H}^{1,1}$ of ${\cal H}^1$. A rank one product projector $\proj{\Psi}$ ($\ket{\Psi} = \bigotimes_{i=1}^N\ket{\Psi^i}$, $\ket{\Psi^i} \in {\cal H}^i$), will be transformed into $Tr_{{\cal Q}^{1,1}}[\proj{\Psi}] = Tr_{{\cal Q}^{1,1}}[\proj{\Psi^1}]\otimes \bigotimes_{i=2}^N\proj{\Psi^i}$, which is a product (in general not pure) state of ${\cal Q}\backslash{\cal Q}^{1,1}$. Therefore $Tr_{{\cal Q}^{1,1}}[\ros]$ is a separable state if $\ros$ is so, and the expression
\begin{equation}
\rho' \equiv Tr_{{\cal Q}^{1,1}}[\rho] = (1+R(\rho))Tr_{{\cal Q}^{1,1}}[\rmes] - R(\rho)Tr_{{\cal Q}^{1,1}}[\rmenys]
\end{equation}
is a local pseudomixture, not necessarily optimal, for the state $\rho'$ of ${\cal Q}\backslash{\cal Q}^{1,1}$ with $t = R(\rho)$. Consequently, $R(Tr_{{\cal Q}^{1,1}}[\rho]) \leq R(\rho)$.$\,\,\,\Box$

Property ($v$) assures that the robustness of the state of a shared system is not an intensive quantity in the number of objects ${\cal Q}$ consists of. Indeed, this follows from the fact that $R(\rho)$ is left unchanged if we give the parties new objects $\tilde{\cal Q}$, provided they are in a separable state $\ros$ and uncorrelated with the objects of ${\cal Q}$. Notice that we need only prove that 
$R(\rho\otimes\ros) \leq R(\rho)$, since Theorem 4 will do the rest.

\vspace{2mm}
{\bf Theorem 5}
\vspace{2mm}

$R(\rho\otimes\ros) \leq R(\rho)$

{\bf Proof:} For $\rho = (1+R(\rho))\rmes - R(\rho)\rmenys$ an optimal local pseudomixture of the state of ${\cal Q}$, the state of ${\cal Q}\bigcup\tilde{\cal Q}$, $\rho\otimes\ros$, admits the following decomposition
\begin{equation}
\rho\otimes\ros = (1+R(\rho))\rmes\otimes\ros-R(\rho)\rmenys\otimes\ros,
\end{equation}
which is a local pseudomixture, not necessarily optimal.$\,\,\,\Box$

{\bf Definition:} Given a realization $\Upsilon \equiv \{\rho_k, p_k\}_{k=1\cdots l}$, we call the quantity $\sum_{k=1}^l p_k R(\rho_k)$ the ({\em average}) {\em robustness} of $\Upsilon$, $R(\Upsilon)$.

 Property ($vi$) refers to the convexity of $R(\rho)$, and it means that the robustness of any realization of $\rho$, $\Upsilon \equiv \{\rho_k, p_k\}_{k=1\cdots l}$, is not smaller than that of $\rho$ itself. It suffices to prove ($vi$) for $l=2$, since $l>2$ can be achieved by iterating this case.

\vspace{2mm}
{\bf Theorem 6}
\vspace{2mm}

$R(p\rho_1 + (1\!\!-\!\!p)\rho_2) \,\leq\, pR(\rho_1)+(1\!\!-\!\!p)R(\rho_2)$, $\, \, \, \, p \in [0, 1]$.

{\bf Proof:} For each $\rho_k$ ($k = 1, 2$) consider an optimal local pseudomixture, say
\begin{equation}
\rho_k = (1\!\!+\!\!R(\rho_k))\,\rho^{+}_{s,k} - R(\rho_k)\rho^{-}_{s,k}.
\end{equation}
 Then $\rho = p\rho_1 + (1\!\!-\!\!p)\rho_2$ can be reexpressed as
\begin{equation}
\rho = (1\!\!+\!t) \,\rmes - t\,\rmenys,
\end{equation}
which is a local pseudomixture, not necessarily optimal, with
\begin{eqnarray}
\rmes &\equiv& \frac{1}{1+t} (p\, (1\!\!+\!\!R(\rho_1))\,\rho^{+}_{s,1} + (1\!\!-\!p)(1\!\!+\!\!R(\rho_2))\,\rho^{+}_{s,2}) \,\,\,\in {\cal S},\\
\rmenys &\equiv& \frac{1}{t} (pR(\rho_1)\,\rho^{-}_{s,1} + (1\!\!-\!p)R(\rho_2)\,\rho^{-}_{s,2}) \,\,\,\in {\cal S},\\
t &\equiv& pR(\rho_1) + (1\!\!-\!\!p)R(\rho_2).
\end{eqnarray}
 Then $R(\rho = p\rho_1 + (1\!\!-\!\!p)\rho_2) \leq t$ by the definition of $R(\rho)$.$\,\,\,\Box$
 
 Let us explain property ($vi$) a bit further. Recall the device $\Sigma$ introduced in section $2.A$. If Alice and Bob are given each one a particle which together are, with equal probability, either in state $\rho_1 = \proj{\Psi_1}$ or $\rho_2 = \proj{\Psi_2}$, with $\ket{\Psi_1} = \ket{1}\otimes\ket{1}$ and $\ket{\Psi_2} =\frac{1}{\sqrt{2}}(\ket{1}\otimes\ket{1}+\ket{2}\otimes\ket{2})$, then if they get them in the separable state $\rho_1 = \proj{\Psi_1}$ together with a message stating this fact, their shared state has null robustness. However, if they destroy the message and forget its content, the new state of the system is $\rho = \frac{1}{2}(\rho_1 + \rho_2)$, which can be checked to be entangled and consequently contains some robustness. This means Alice and Bob have increased the robustness of their system by acting locally. Notice however that if the process is repeated many times (each repetition consisting of first getting a couple of particles along with a message stating their global state, and then destroying the message and forgetting its content), on average the robustness of the freshly obtained couples is $\frac{1}{2}R(\rho_1) +\frac{1}{2}R(\rho_2)$, whereas the robustness of the state the couples finally end up in is $R(\rho) \leq \frac{1}{2}(R(\rho_1) + R(\rho_2))$. Therefore $(vi)$ states that one cannot, in a statistical sense, increase the robustness of a shared state by mixing.

 Let us discuss now property $(vii)$, which assures that the output of a local measurement on $\rho$ is a realization $\Upsilon = \{\rho_k, p_k\}_{k=1\cdots l}$ (of the {\em averaged} final state $\rho_f\equiv\sum_k p_k\rho_k$) that cannot have more robustness than $\rho$, so that the robustness of a system ${\cal Q}$ cannot be increased, on average, by performing a local measurement on it. 
 
 Although property $(vii)$ refers to an incomplete local von Neumann measurement (that is, one implemented by a set of orthogonal product projectors, not necessarily of rank one but wich correspond to a resolution of the identity), we will prove it for local measurements of the most general nature. In addition to being incomplete, they may include the temporary use of ancillas (local POVM's) and classical communication between the parties, and contemplate conditional rejection of the system depending on the output. A general local measurement is implemented by a set $\{A_k\}_{k=1\cdots l}$ of product operators that satisfy $0 \leq \sum_k A_k^{\dagger}A_{k} \leq I$. As a result of such a measurement the state of the system becomes, with probability $p_k = Tr[A_k\rho A_k^{\dagger}]$, $\rho_k = \frac{A_k\rho A_k^{\dagger}}{Tr[A_k\rho A_k^{\dagger}]}$. Notice that $\sum_k p_k \leq 1$. Thus the realization $\Upsilon = \{\rho_k, p_k\}_{k=1\cdots l}$ and the averaged final state $\rho_f$ are in general unnormalized.

\vspace{2mm}
{\bf Theorem 7}
\vspace{2mm}

 If the (unnormalized) realization $\Upsilon = \{\rho_k, p_k\}_{k=1\cdots l}$ describes the potential final states of a general local measurement performed on ${\cal Q}$ in the state $\rho$, then $R(\rho) \geq R(\Upsilon)$.

{\bf Proof:} One can check that for $\ros \in {\cal S}$, any resulting state $\rho_{s,k}$ is separable as well (as it is to be expected, otherwise we would get some entanglement out of a separable state, even in a statistical sense). Then, for $\rho = (1+R(\rho))\rmes - R(\rho)\rmenys$ an optimal local pseudomixture of $\rho$, we can write

\begin{equation}
\rho_k = \frac{1}{Tr[A_k\rho A_k^{\dagger}]}((1+R(\rho))A_k\rmes A_k^{\dagger} - R(\rho)A_k\rmenys A_k^{\dagger}),
\end{equation}
that implies that if $\rho_k = (1+R(\rho_k))\rho^+_{s,k} - R(\rho_k)\rho^-_{s,k}$ is an optimal local pseudomixture, then $R(\rho_k) \leq R(\rho) \frac{Tr[A_k\rmenys A_k^{\dagger}]}{Tr[A_k\rho A_k^{\dagger}]}$. Therefore $R(\Upsilon) = \sum_k p_k R(\rho_k) = \sum_k Tr[A_k\rho A_k^{\dagger}] R(\rho_k) \leq $ $R(\rho) Tr[\sum_k A_k^{\dagger}A_k\rmenys] \leq R(\rho) Tr[I\rmenys] = R(\rho)$.$\,\,\,\Box$

\vspace{4mm}

  We want to stress here that properties ($i$), ($iii$)-($vii$) must be satisfied by any magnitude $\mu(\rho)$ consistent with the fundamental law of quantum information processing \cite{VP2} \cite{PopRoh}, that is, non-increasing under local actions of the parties, which are allowed to communicate classically. Properties ($iii$)-($vii$) must be obviously satisfied (for property ($vi$), see discussion in section 2.A), whereas property ($i$) is also necessary, but it can be proved to follow from properties ($iii$)-($v$). Notice that properties ($iii$)-($vii$), which we claim to be a set of {\sl necessary and sufficient properties a magnitude $\mu$ has to fulfill in order to be consistent with the fundamental law of quantum information processing}, do not mention the fact that the parties can share information using a classical channel. The reason for this is that the use of classical communication simply allows for a wise selection of a local action conditioned to the result of previous local measurements, each of these local actions not increasing, on average, the magnitude $\mu$. Properties ($iii$)-($vii$) are sufficient because any operation the parties can perform locally on the local subsystems can be decomposed into elementary steps taken into account in ($iii$)-($vii$).


\vspace{4mm}
 
 Finally, property ($viii$) is a very weak version of a composition law \footnote{
Additivity of the robustness, that is $R(\rho_x\otimes\rho_y)=R(\rho_x)+R(\rho_y)$, would be a particular form of a composition law. We already know, however, that the robustness is not an additive quantity, as will be shown elsewhere, though a function of it could well be additive.}. 
Consider a state of the form $\rho_x\otimes\rho_y$, where, for instance, $\rho_x$ may be the global state of a non-local system which consist of four particles shared by Alice, Bob and Claire, whereas $\rho_y$ may be that of other five particles shared by Alice and Denise. The lack of correlations between $\rho_x$ and $\rho_y$ allows on one hand Alice, Bob and Claire to mix $\rho_x$ with a separable state $\rho_{s,x}^-$ with weight $R(\rho_x)$ and on the other Alice and Denise to mix $\rho_y$ with a separable state $\rho_{s,y}^-$ with weight $R(\rho_y)$. These operations transform $\rho_x\otimes\rho_y$ into $\rho_{s,x}^+\otimes\rho_{s,y}^+$, a separable state different from any separable state associated with $R(\rho_x\otimes\rho_y)$. Whether $R(\rho_x\otimes\rho_y)$ is determined or not by $R(\rho_x)$ and $R(\rho_y)$, we will now show that the knowledge of $R(\rho_x)$ and $R(\rho_y)$ leads to bounds on $R(\rho_x\otimes\rho_y)$, which is what property ($viii$) announces.
 These bounds are
\begin{equation}
\max(R(\rho_x), R(\rho_y))\,\,\, \leq \,\,R(\rho_x\otimes\rho_y)\,\, \leq \,\,\,R(\rho_x)+R(\rho_y)+2R(\rho_x)R(\rho_y),
\end{equation}
and can be obtained as follows: the lower bound results from taking the partial trace over the Hilbert space of either $\rho_x$ or $\rho_y$ in an optimal local pseudomixture for $\rho_x \otimes \rho_y$, and is a consequence of property ($iv$), whereas to deduce the upper bound one needs to take into consideration the tensor product of two optimal local pseudomixtures for the two shared states $\rho_x$ and $\rho_y$ ($x \equiv R(\rho_x)$, $y \equiv R(\rho_y)$), which is a local pseudomixture for $\rho_x \otimes \rho_y$, not necessarily optimal, with weight $t = x + y +2xy$,

\begin{eqnarray}
\rho_x \otimes \rho_y = [(1+x)\rho^+_{s,x} - x\rho^-_{s,x}]\otimes[(1+y)\rho^+_{s,y} - y\rho^-_{s,y}] \nonumber\\
= (1+x)(1+y)\;\rho^+_{s,x}\!\!\otimes\!\rho^+_{s,y} +xy\;\rho^-_{s,x}\!\!\otimes\!\rho^-_{s,y} \nonumber\\
-  \{ x(1+y) \rho^-_{s,x}\!\!\otimes\!\rho^+_{s,y} + (1+x)y\;\rho^+_{s,x}\!\!\otimes\!\rho^-_{s,y}\}.
\end{eqnarray}

\vspace{2mm}
{\bf 3.C Numerical computations and convexity}
\vspace{2mm}

 We end the exposition of general properties of the robustness $R(\rho)$ by mentioning a property of the relative robustness $R(\rho||\ros)$ which is most valuable for the numerical computation of the absolute robustness of a state $\rho$, $R(\rho)$, namely that $R(\rho ||\ros)$ is a convex function of $\ros$.

 Indeed, if 
\begin{equation}
\rho = (1+R_k)\,\rho_{s,k}^+-R_k\,\rho_{s,k} \equiv [k],\,\,\,\,(k\, =\, 1,2) 
\end{equation}
is the local pseudomixture for $\rho$ that, involving the separable state $\rho_{s,k}$, has minimum non-negative weight $R_k \equiv R(\rho||\rho_{s,k})$ (cf. Eq.(\ref{relative})), then the convex combination 
\begin{equation}
\frac{1}{\frac{p}{R_1}+\frac{1-p}{R_2}}\,(\frac{p}{R_1}[1]+\frac{1-p}{R_2}[2]),
\end{equation}
is another local pseudomixture for $\rho$, involving $\ros \equiv p\rho_{s,1} + (1\!-\!p)\rho_{s,2}$, with weight $t = \frac{1}{\frac{p}{R_1}+\frac{1-p}{R_2}}$.
Since $(\frac{p}{R_1}+\frac{1-p}{R_2})(pR_1+(1\!-\!p)R_2) = p^2 + (1-p)^2 +(\frac{R_1}{R_2}+\frac{R_2}{R_1})p(1-p) \geq 1$, it follows that 
\begin{equation}
R(\rho ||\ros) \leq t \leq pR(\rho||\rho_{s,1}) + (1-p)R(\rho||\rho_{s,2}).
\end{equation}

 This means that if $R(\rho)$ is computed by searching in the set of separable states ${\cal S}$ for the absolute minimum of $R(\rho||\ros)$ as a function of $\ros$, then the search can finish as soon as a local minimum is found, for any local minimum of $R(\rho ||\ros)$ is also the absolute one. We will use this fact in section 4.A to explain an efficient way of numerically computing $R(\rho)$ for states of the two simplest binary composite systems.

\vspace{3mm} 
{\bf 4. ROBUSTNESS AND RANDOM ROBUSTNESS OF BINARY COMPOSITE SYSTEMS.}

\vspace{2mm} 

 So far all our considerations have referred to composite systems with an unrestricted number of parties $N$. We move to consider in what follows a composite system ${\cal Q}$ shared by two parties, Alice and Bob, so that from now on $N=2$. Recall that, as before, ${\cal H}^i \cong {\cal C}^{n_i}$ is the Hilbert space of all the physical objects party $i$ can act locally on.

\vspace{3mm} 
{\bf 4.A Robustness of binary composite systems.}
\vspace{2mm} 
 
 We present here a list of bounds and exact results concerning the robustness of states of a binary system. A method for numerically computing this quantity for the two simplest binary systems is also discussed. These results make the robustness of states of binary systems useful as an entanglement magnitude. And thus, for instance, from its expression for pure states one can see that robustness, together with the entropy of entanglement, can be used to completely characterize the entanglement of pure states of a two-qutrit system (see Introduction).

\vspace{3mm} 
{\bf Robustness of pure states of binary composite systems.}
\vspace{2mm} 

 It turns out that for binary systems with Hilbert space ${\cal C}^m \!\otimes{\cal C}^m$ a set of $m-1$ ordered non-negative parameters $\{a_i\}_{i=1,...m-1}$ suffices to completely specify any element of the set of locally inequivalent pure states\footnote{
In general, any two states $\rho_1$ and $\rho_2$  are said to be locally equivalent if they are related by a unitary product transformation, i.e. if $\rho_1 = U_L\rho_2 U_L^\dagger$.},
\begin{equation}
\frac{({\cal C}^m \!\otimes{\cal C}^m \,\backslash \,\{\ket{0}\})/R^+}{{\cal U}(m)\, \mbox{x}\,{\cal U}(m)}
\end{equation}
(that is the space of the orbits, in the subset of normalised elements of the complex vector space ${\cal C}^m \!\otimes{\cal C}^m$, of the action of all unitary product transformations). This set $\{a_i\}$ can easily be obtained for any normalized vector $\ket{\Psi}$ from its ordered Schmidt decomposition, 
\begin{equation}
\ket{\Psi} = \sum_{i=1}^m a_i \ket{i}\!\otimes\!\ket{i}; \,\,\, a_i \geq a_{i+1} \geq 0, \sum_{i=1}^m a_i^2 = 1,
\end{equation}
after excluding $a_m$. It will be more convenient, however, to keep all $m$ coefficients. Then, in terms of $\{a_i\}$, the robustness $R$ of the pure state $\Psi$ is:
\begin{equation}
R(\Psi(\{a_i\})) =  (\sum_{i=1}^{m} a_i)^2 - 1.
\label{ext}
\end{equation}

 This result is proved in Appendix B, and indicates how $R(\Psi)$ can be systematically computed: given a rank one projector corresponding to a pure state, $\rho = \proj{\Psi}$, one needs only to perform a partial trace over any of the two parties, and get the eigenvalues of the remaining matrix. These eigenvalues are $a_i^2$, so that the sum of their square roots will immediately lead to $R(\Psi)$.
 
Notice that the sets
\begin{equation}
\frac{{\cal C}^m \!\otimes{\cal C}^m}{{\cal U}(m) \, \mbox{x}\,{\cal U}(m)} \,\,\,\,\mbox{and}\,\,\,\,  \frac{{\cal C}^{n_1} \!\otimes{\cal C}^{n_2}}{{\cal U}(n_1) \, \mbox{x}\,{\cal U}(n_2)}
\end{equation}
for any $n_1, n_2$ satisfying $m = \min(n_1,n_2)$ are equivalent (since the Schmidt decomposition of $\ket{\Psi} \in {\cal C}^{n_1} \!\otimes{\cal C}^{n_2}$ contains at most $m$ terms), so that Eq.(\ref{ext}) is also valid for any state in ${\cal C}^{n_1} \!\otimes{\cal C}^{n_2}$ if $m = \min(n_1,n_2)$.

 One can check that, as previously announced, for $m = 3$ the entropy of entanglement $E(\Psi)$ given in Eq. (\ref{entropy}) and the robustness $R(\Psi)$ are independent functions of the two greatest Schmidt coefficients $a_1$ and $a_2$, and that there is a one-to-one correspondence between $(a_1, a_2)$ and $(E, R)$, so that $(E,R)$ can be used to label unambiguously the elements of the set of locally inequivalent pure states of a two-qutrit system, and therefore suffices to completely characterize their entanglement.

\vspace{3mm} 
{\bf Bounds for the robustness of mixed states of binary composite systems.}
\vspace{2mm} 

 It can be proved (see Appendix C) that for any state of a binary system the following inequalities hold:

\begin{equation}
|\min(\{\frac{\lambda_j}{a_{j,1}^2}\},0)| \leq R(\rho) \leq \min(\tilde{m}-1,R(\tilde{\rho}||\frac{1}{\tilde n}\tilde{I}))
\label{robmix}
\end{equation}
where $\lambda_j$ is the $j^{th}$ negative eigenvalue of $\rho^{T_B}$,\footnote{
$\rho^{T_B}$ is the partial transposed of $\rho$ with respect the party B (which has the same spectrum as $\rho^{T_A}$, its eigenvectors also having the same Schmidt coefficients).
}
 $a_{j,1}$ is the biggest coefficient of the Schmidt decomposition of the eigenvector corresponding to $\lambda_j$, $\tilde{n}$ is the rank of $\rho^A\otimes\rho^B \equiv Tr_B[\rho]\otimes Tr_A[\rho]$ (i.e., the dimension of the minimum product space $\tilde{\cal H} \subseteq {\cal H}$ such that $\rho$ is entirely supported in it), $\tilde{m} = \min($rank$[\rho^A],$ rank$[\rho^B])$ and $\tilde{\rho}$ and $\tilde{I}$ are the restrictions of $\rho$ and $I$ to $\tilde{\cal H}$. 

\vspace{3mm} 
{\bf Robustness of a two qubit system.}
\vspace{2mm} 

 For the simplest binary composite system, the ${\cal C}^2\otimes {\cal C}^2$ case, we present simpler bounds for the robustness of a general mixed state and an exact result for a class of mixed states, which includes all Werner states. These results are proved in Appendix C.

 First, for $\lambda$ the negative eigenvalue of $\rho^{T_B}$ and $\ket{n}$ its corresponding eigenvector, with $\ket{n} = \cos{\theta}\ket{1}\!\otimes\!\ket{1} + \sin{\theta}\ket{2}\!\otimes\!\ket{2}\,$ ($\theta \in [0,\frac{\pi}{4}]$) its ordered Schmidt decomposition, the following inequalities hold for any state $\rho$:

\begin{equation}
\frac{|\lambda|}{\cos^2\theta} \leq R(\rho) \leq  2|\lambda |,
\end{equation}
which in particular means that whenever $\cos^2\theta = \frac{1}{2}$, $R(\rho) = 2 |\lambda |$. The lower bound corresponds to Eq.(\ref{robmix}), and the upper bound can be seen to be an improvement on that in Eq.(\ref{robmix}) by taking into account the result in Eq.(\ref{randc2c3}) and that $|\lambda| \leq \frac{1}{2}$ \cite{volume}, $\tilde{m} = 2$ for any entangled $\rho$.

Another upper bound for the robustness comes from the fact that for pure states of ${\cal C}^2\otimes {\cal C}^2$ the concurrence $C(\Psi)$ (see  \cite{Wootters}) equals the robustness, and it reads:

\begin{equation}
R(\rho) \leq C(\rho),
\end{equation}
 where $C(\rho)$ was explicitly computed for any state of this system in  \cite{Wootters}. 

Finally, we have computed the robustness for a family of mixed states: consider the rank one projector $\proj{\theta}$, where $\ket{\theta} \equiv \cos{\theta}\ket{1}\!\otimes\!\ket{1} + \sin{\theta}\ket{2}\!\otimes\!\ket{2}$, $\theta \in [0,\frac{\pi}{4}]$, and the (separable) diagonal state

\begin{equation}
\rho_D \equiv \left( \begin{array}{cccc}
q_1 & 0 & 0 & 0  \\
0 & \frac{q_2}{2} & 0 & 0 \\
0 & 0 & \frac{q_2}{2} & 0 \\
0 & 0 & 0 & q_3 
\end{array} \right), \,\,\, q_i \geq 0,\,\,\, \sum_i^3 q_i=1;
\end{equation}
then, for any $0 \leq p \leq 1$, the state $\rho \equiv p\rho_D + (1\!-\!p)\proj{\theta}$ has robustness 
\begin{equation}
R(\rho) =  \left\{ \begin{array}{cl}
0  & \mbox{if}\,\,\,\, \rho^{T_B} \geq 0  \\
(1\!-\!p)\sin{2\theta}-pq_2 & \mbox{otherwise.}
 \end{array}\right. 
\end{equation}
 
 A Werner state with fidelity $F$ \cite{Werner} is locally equivalent to the $\rho$ resulting from taking $q_1=q_3=\frac{q_2}{2}= \frac{1}{4}$, $\theta = \frac{\pi}{4}$ and $p=4\frac{1-F}{3}$, and in terms of its fidelity we have $R(\rho) = 2F-1$ for entangled Werner states, that is for Werner states with fidelity $F > \frac{1}{2}$.

\vspace{2mm} 
{\bf Numerical computation of the robustness for mixed states of two qubits and of a qubit-qutrit system.}
\vspace{2mm} 

 In ${\cal C}^2\otimes {\cal C}^2$ and ${\cal C}^2\otimes {\cal C}^3$ one can easily check whether a state $\rho$ is separable by computing the eigenvalues of $\rho^{T_B}$ and seeing whether they all are non-negative, since for these systems $\rho \in {\cal S} \Leftrightarrow \rho^{T_B} \geq 0$ \cite{Peres},\cite{Horo2}. Therefore given a $\rho$ which is known to be entangled, one can choose a separable state $\ros$ and compute $R(\rho||\ros)$ by requiring that $s$ in Eq.(\ref{relative}) be minimum with $\rho(s)^{T_B}\geq 0$. Consequently to find $R(\rho)$ one can perform, say, a conditional random walk, in the 16 (or 36) dimensional real vector space of hermitian 4x4 (or 6x6) matrices $s\ros$, searching for the minimum of its trace $s$, requiring
\begin{eqnarray}
\ros \geq 0, \label{c1}\\
\ros^{T_B} \geq 0,\label{c2}\\
(\rho + s\ros)^{T_B} \geq 0, \label{c3}
\end{eqnarray}
and that at each step $s$ diminishes.
 Conditions (\ref{c1}) and (\ref{c2}) assure that $\ros$ is a separable state, and then condition (\ref{c3}) assures that $\frac{1}{1+s}(\rho + s\ros)$ is also separable. For each $s\ros$ satisfying conditions (\ref{c1}), (\ref{c2}) and (\ref{c3}), $s$ is greater than or equal to $R(\rho||\ros)$, and from the convexity of this function (see section 3.C) we know the search will finish as soon as a local minimum is reached for $s$, for it is the global one.

 In ${\cal C}^2\otimes {\cal C}^2$ the effectiveness of this method is notably enhanced by the fact that, as a consequence of some results of  \cite{STV}, a state $\rho$ of this system is entangled if, and only if, $\det \rho^{T_B} < 0$, that is,
\begin{equation}
\rho \in {\cal S} \Leftrightarrow \det \rho^{T_B} \geq 0.
\end{equation} 
 Then, whereas the eigenvalues of $\ros$ must be computed to check constraint (\ref{c1}), for constraints (\ref{c2}) and (\ref{c3}) one only needs to compute the determinant of $\ros^{T_B}$ and that of $(\rho+s\ros)^{T_B}$.

\vspace{3mm} 
{\bf 4.B Random robustness of binary composite systems.}
\vspace{2mm} 

 The random robustness of binary composite systems, that we shall compute exactly for pure states of a system ${\cal C}^{n_1}\otimes {\cal C}^{n_2}$ and for any state of the systems ${\cal C}^2\otimes {\cal C}^2$ and ${\cal C}^2\otimes {\cal C}^3$, and for which we will present lower and upper bounds for any state in any system, is a quantity that will be very useful for two different purposes. We proved in section 2.B that any state of any composite system can be expressed in terms of two separable states, $\rmes$ and $\rmenys$, and a non-negative number $t$, i.e. as a local pseudomixture. Moreover, we provided an explicit offhand example of local pseudomixture for any state $\rho$. However we didn't prove this last result, and this is what we will do with the help of the random robustness of mixed states. On the other hand this quantity will allow us to obtain an explicit lower bound for the volume of separable states of a generic composite system in section 5.

\vspace{3mm} 
{\bf Random robustness of pure states of binary composite systems.}
\vspace{2mm} 

 Given a pure state $\Psi$ of a binary system ${\cal C}^{n_1}\otimes {\cal C}^{n_2}$ with ordered non-local parameters $\{a_i\}_{i=1,m}$ ($m = \min(n_1, n_2))$, its random robustness is (Appendix B)

\begin{equation}
R(\Psi||\frac{1}{n_1n_2}I) = n_1n_2a_1a_2,
\end{equation}
which manifestly depends not only on the two largest coefficients $a_1$ and $a_2$ (that is, on the state itself), but also on the dimension $n=n_1n_2$ of the Hilbert space of the system (cf. property ($i$) of $R(\rho)$). Notice that for any dimensions the most robust pure state, as far as white noise is concerned, has $a_1=a_2 = \frac{1}{\sqrt{2}}$, and thus is locally equivalent to a singlet state in a ${\cal C}^2\otimes {\cal C}^2$ product subspace of ${\cal C}^{n_1}\otimes {\cal C}^{n_2}$.

\vspace{3mm} 
{\bf Bounds for the random robustness of mixed states of binary composite systems.}
\vspace{2mm} 

 For any $\rho$ of a binary system with Hilbert space ${\cal C}^{n_1}\otimes {\cal C}^{n_2}$ of dimension $n=n_1n_2$, and for $\lambda$ the smallest eigenvalue of $\rho^{T_B}$, the following bounds hold (Appendix C):

\begin{equation}
n|\min(\lambda,0)| \,\, \leq \,\, R(\rho||\frac{1}{n}I) \,\, \leq \,\, \frac{n}{2}
\label{rrmixed}
\end{equation}

 The upper bound is of some interest, for it indicates how any state of a binary system can be offhand explicitly written in terms of a local pseudomixture, and it can be generalized to the $N$-party case, where it reads

\begin{equation}
R(\rho||\frac{1}{n}I) \,\, \leq \,\, (1+\frac{n}{2})^{N-1} -1,
\end{equation}
as it was already mentioned at the end of section 2.B.

\vspace{3mm} 
{\bf Random robustness of a two-qubit system and of a qubit-qutrit system.}
\vspace{2mm} 

 Because in ${\cal C}^2\otimes {\cal C}^2$ and ${\cal C}^2\otimes {\cal C}^3$ the condition $\rho^{T_B} \geq 0$ is not only necessary but also sufficient for $\rho$ to be separable \cite{Peres},\cite{Horo2}, the lower bound in Eq.(\ref{rrmixed}), which was based on this condition, becomes an equality:

\begin{equation}
R(\rho\,||\frac{1}{n}I) = n|\min(\lambda,0)|.
\label{randc2c3}
\end{equation}

\vspace{4mm}
{\bf 5. APPLICATION: EXPLICIT LOWER BOUND FOR THE VOLUME OF SEPARABLE STATES.}
\vspace{4mm} 

 In  \cite{volume} the space of states ${\cal T}$ was endowed with a measure, for which it was proved that the volume of the set of separable states ${\cal S}$ was non-zero compared to that of the whole set of states ${\cal T}$. We will next give an alternative proof of this result by computing an explicit lower bound for this volume. Following the proposal in  \cite{volume}, the set of states of a generic system ${\cal Q}$ can be viewed as a Cartesian product of two sets:
\begin{equation}
{\cal T} \sim {\cal P}\mbox{{\sf x}}\Delta
\end{equation}
where ${\cal P}$ is the set of complete families $\{P_k\}_{1\cdots n}$ of orthogonal rank one projectors (i.e. $\sum_{k=1}^n P_k = I, \,\,Tr[P_kP_{k'}] = \delta_{k,k'}, \,\, P_k^2 = P_k$), and $\Delta$ is the convex subset of ${\cal R}^n$ generated by all possible convex combinations of the points $x_i \in {\cal R}^n$, $x_i \equiv ( 0, ..., 0, 1_i, 0,..., 0),\,\,\, i=1\cdots n.$ (that is, $\Delta$ is the convex hull generated by $\{x_i\}_{i=1\cdots n}$ and thus a subset of the $n-1$ dimensional hyperplane which contains $\{x_i\}$). For $\nu$ the measure induced on ${\cal P}$ by the Haar measure on the unitary group $U(n)$ and ${\cal L}_{n-1}$ the Lebesgue measure induced on $\Delta \subset {\cal R}^{n-1}$, it was argued in  \cite{volume} that a natural measure on ${\cal T}$ is $\mu = \nu$ {\sf x} ${\cal L}_{n-1}$. We have then found the following lower bound for the ratio of the volume of the sets ${\cal S}$ and ${\cal T}$ of a $N$-party system with $n$-dimensional Hilbert space:

\begin{equation}
\frac{V({\cal S})}{V({\cal T})} \geq \left(\frac{1}{1+\frac{n}{2}}\right)^{(n\!-\!1)(N\!-\!1)},
\end{equation}
which indeed confirms that the volume of separable states is non-zero for any finite $n$.

{\bf Proof:} 
Consider the function
\begin{equation}
\Theta(\{P_k\},\{\Lambda_k\}) \equiv \left\{ \begin{array} {cl}
1 & \mbox{if}\,\, \sum_{k=1}^n \Lambda_k P_k \,\,\in {\cal S}\\
0 & \mbox{otherwise,}
\end{array} \right.
\end{equation}
where $\{\Lambda_k\} \in \Delta$. Then the ratio of the volumes $V({\cal S})$ and $V({\cal T})$ is, with the proposed measure $\mu = \nu$ {\sf x} ${\cal L}_{n-1}$ on ${\cal T} = {\cal P}$ {\sf x} $\Delta$,
 
\begin{equation}
\frac{V({\cal S})}{V({\cal T})} =\frac{\int_{U(n)}dU\int_{\Delta}d\Delta\,\Theta(\{P_k\},\{\Lambda_k\})}{\int_{U(n)}dU\int_{\Delta}d\Delta}.
\end{equation}

 Consider now another function $\Xi(\{P_k\},\{\Lambda_k\}) \leq \Theta(\{P_k\},\{\Lambda_k\})$. Then

\begin{equation}
\frac{V({\cal S})}{V({\cal T})} \geq \frac{\int_{U(n)}dU\int_{\Delta}d\Delta\,\Xi(\{P_k\},\{\Lambda_k\})}{\int_{U(n)}dU\int_{\Delta}d\Delta}.
\label{quocient}
\end{equation}
 If one can choose this function $\Xi$ to be independent of $\{P_k\}$, then the integral over the unitary group in the numerator of Eq.(\ref{quocient}) will factor out and will be cancelled by that in the denominator. As we will argue, the following one does the job:
\begin{equation}
\Xi(\{\Lambda_k\}) \equiv \left\{ \begin{array} {cl}
1 & \mbox{if}\,\, \{\Lambda_k\} \in \Delta_p\\
0 & \mbox{otherwise,}
\end{array} \right.
\end{equation}
where $\Delta_p \equiv \mbox{convexhull}\{y_i \in {\cal R}^n; y_i = px_i+(1\!-\!p)z_I,\,\,\, i=1\cdots n\}$, with $z_I\equiv (\frac{1}{n},...,\frac{1}{n})$ and $p=\left(\frac{1}{1+\frac{n}{2}}\right)^{N-1}$. Then one can see that, since the simplex $\Delta_p$ has edges $p$ times smaller than $\Delta$,
  
 \begin{equation}
\int_{\Delta}d\Delta\,\Xi(\{\Lambda_k\}) =\int_{\Delta_p}d\Delta= p^{n-1} \int_{\Delta}d\Delta,
\end{equation}
from where the lower bound easily follows.

 To see that any state $\sum_{k=1}^n \Lambda_k P_k$ is separable for any family $\{P_k\}$ provided that $\{\Lambda_k\} \in \Delta_p$ (that is, to see that $\Theta \geq \Xi$), one can ressort to the upper bound for the random robustness Eq.(\ref{cota}) computed at the end of Appendix C. Since $R(\rho||\frac{1}{n}I) \leq \left(1+\frac{n}{2}\right)^{N\!-\!1} - 1 \equiv \tilde{t}$, we find that a $p$, independent of $\rho$, such that 
\begin{equation}
p\rho + (1\!-\!p)\frac{1}{n}I
\end{equation}
belongs to the set of separable states ${\cal S}$, is $p\equiv\frac{1}{1+\tilde{t}} = \left(\frac{1}{1+\frac{n}{2}}\right)^{N\!-\!1}$. Each point $\{\Lambda_k\} \in \Delta_{p}$ has components $\Lambda_k = q_kp+\frac{(1-p)}{n}$ for some $q_k \geq 0$ such that $\sum_{k=1}^n q_k = 1$. Then
\begin{equation}
\sum_{k=1}^n \Lambda_kP_k = p\sum_{k=1}^n q_kP_k + (1\!-\!p)\sum_{k=1}^n\frac{P_k}{n} = \sum_{k=1}^n q_k(pP_k + (1\!-\!p))\frac{1}{n}I),
\end{equation}
which is a convex combination $ \sum_{k=1}^n q_k \rho_{s,k}$ of separable states $\rho_{s,k} \equiv pP_k + (1\!-\!p))\frac{1}{n}I$, and therefore is also separable.

\vspace{4mm}
{\bf Acknowledgements.}
\vspace{4mm}

 G.V. would like to thank the members of the Department of Physics, University of Wales, Swansea, for their hospitality during his stay there, where part of this paper was elaborated. He is specially grateful to Mark Emmett, Ben White, Alex Dougall and Ho Joon Lee. G.V. also acknowledges a CIRIT grant 1997FI-00068 PG. R.T. acknowledges financial support from CICYT, contract AEN95-0590, and from CIRIT, contract 1996GR00066.

\vspace{4mm}
{\bf Appendix A: Notation.}
\vspace{4mm}

 Entanglement appears in composite systems, where divisions - and yet subdivisions - into constituent parts easily proliferate. These may imply working with a multitude of Hilbert spaces which, together with having to deal with different types of states, easily leads to confusion. We have chosen to label symbols such as ${\cal Q}$, ${\cal H}$, $\rho$ (standing for physical systems, Hilbert spaces, states, ...) with a superindex to refer to a specific local subsystem, the parties being called after the name of a physicist - Alice, Bob, ... - following the tradition. On the other hand subindices will denote different elements of a collection of states. 

\begin{figure}
 \epsfysize=6.0cm
 \epsffile{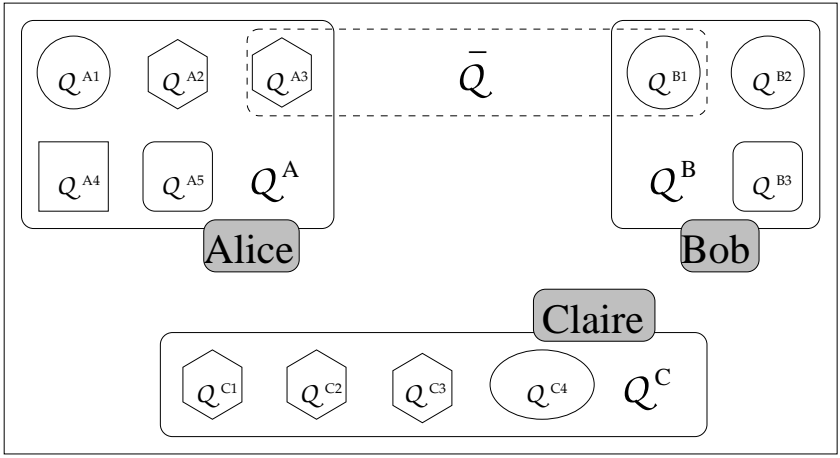}
 \caption{Example of a composite system shared by parties. Twelve local partial subsystems of five different types, and thus not all identical, are grouped together into three local subsystems, according to which physicist or party - Alice, Bob or Claire - can act on them. One can also consider non-local subsystems, such as $\bar{\cal Q} \equiv {\cal Q}^{A3} \bigcup{\cal Q}^{B1} $, which involve partial subsystems belonging to different local subsystems.}
 \label{fig3}
\end{figure}

 The following list contains some of the symbols we have used, along with a short explanation of their meaning
. In some cases we indicate how they are related to each other. See also the example in Figure \ref{fig3}.

\begin{eqnarray}
\begin{array}{rl}
{\cal Q}: & \mbox{physical system, composed of $N$ local subsystems.} \\
{\cal H}: & \mbox{Hilbert space of ${\cal Q}$, of dimension $n$.} \\ 
\Psi, \Phi,... & \mbox{pure states of ${\cal Q}$.} \\
\rho : & \mbox{mixed state of ${\cal Q}$, or exceptionally of a non-local subsystem of ${\cal Q}$.} \\
&\\
{\cal Q}^i: & \mbox{local subsystem $i$ ($i=1\cdots N$), i.e. subsystem where party $i$ can act on} \\
        & \mbox{without further ado (the index $i$ will often be a capital letter instead of a number,}\\
        & \mbox{that is $i = A,B,C,...$}).\\
{\cal H}^i: & \mbox{Hilbert space of ${\cal Q}^i$, of dimesion $n_i$.} \\
\Psi^i,\,\rho^i : & \mbox{states of ${\cal Q}^i$.} \\
& \bigcup_{i=1}^N {\cal Q}^i = {\cal Q}, \hspace{3mm} \bigotimes_{i=1}^N {\cal H}^i = {\cal H} \,\,\,(\prod_{i=1}^N n_i = n).\\
&\\
{\cal Q}^{i,j}: & \mbox{local partial subsystem or part $j$ of the local subsystem $i$, $j=1\cdots N_i$.} \\
{\cal H}^{i,j}: & \mbox{Hilbert space of ${\cal Q}^{i,j}$}. \\
& \bigcup_{j=1}^{N_i} {\cal Q}^{i,j} = {\cal Q}^i, \hspace{3mm} \bigotimes_{j=1}^{N_i} {\cal H}^{i,j} = {\cal H}^i. \\
{\cal Q}\backslash \tilde{\cal Q}: & \mbox{system obtained from ${\cal Q}$ by dismissing a (local or non-local) subsystem $\tilde{\cal Q}$.}\\
&\\
\rho_k, t_k...: & \mbox{element $k$ of a collection of states, weights, ... (typically $k=1\cdots l$)}.\\
&\\
{\cal T}: &  \mbox{set of states.}\\
{\cal S}: &  \mbox{set of separable states.}\\
\rho_s: &  \mbox{separable state, i.e. } \,\rho_s \in {\cal S}. \end{array}
\nonumber 
\end{eqnarray}
 
\renewcommand{\theequation}{B.\arabic{equation}}
\setcounter{equation}{0}

\vspace{4mm}
{\bf Appendix B: Robustness and random robustness of pure states of binary systems. }
\vspace{4mm} 

 Proving that the robustness of any pure state $\Psi$ in ${\cal C}^m \!\otimes{\cal C}^m$ is
\begin{equation}
R(\Psi(\{a_i\})) =  (\sum_{i=1}^m a_i)^2 - 1
\label{purs}
\end{equation}
will take two steps. In the first step a local pseudomixture $\{\rmes, \rmenys, t\}$, such that $t=(\sum_i a_i)^2 - 1$, will be explicitly constructed for $\Psi$. In the second one it will be proved that this local pseudomixture $\{\rmes, \rmenys\, t\}$ is optimal, so that $t(\Psi; \rmes, \rmenys)$ is $R(\Psi)$. Then from Eq.(\ref{purs}) it will be easy to obtain the random robustness of $\Psi$,
\begin{equation}
R(\,\Psi(\{a_i\})\, || \frac{1}{n}I ) = na_1a_2, 
\end{equation}
To further illustrate the issue, the case ${\cal C}^2 \!\otimes{\cal C}^2$ will be treated in more detail.
\vspace{2mm} 

 {\bf B.1} Consider, thus, a pure state $\Psi \in {\cal C}^m \!\otimes{\cal C}^m$, and its ordered Schmidt decomposition:
\begin{equation}
\ket{\Psi} = \sum_{i=1}^m a_i \ket{ii}; \,\,\,\,\,\, a_i \geq a_{i+1} \geq 0,\,\,\, \sum_{i=1}^m a_i^2 = 1, 
\end{equation}
where, from now on, $\ket{ef} \equiv \ket{e}\!\otimes\ket{f} \in {\cal C}^m \!\otimes{\cal C}^m$ and, unless otherwise specified, $\{\ket{i}\}_{i=1\cdots m}$ is an orthonormal basis in ${\cal C}^m$.
 We are interested in statistically mixing $\Psi$ with pure product states in such a way that the final mixture is separable and the statistical weight of the separable part is minimal. Let us define $R \equiv \sum_{i \neq j} a_i a_j = (\sum_i a_i)^2 - 1$ and also
\begin{eqnarray}
\rmenys &\equiv& \frac{1}{R} \sum_{i \neq j} a_i a_j \proj{ij} \label{rmenys} \\
\rmes &\equiv& \frac{1}{1+R}(\proj{\Psi} + R\rmenys). \label{rmes}
\end{eqnarray}
 Notice that $\rmenys$ is a separable state by construction, since it has been built as a convex combination of projectors onto product vectors $\ket{ij}$. Next it will be shown that $\rmes$ is a separable state as well.

 Consider the following convex combination:

\begin{equation}
\rho_s \equiv \frac{1}{\alpha_{m+1}} \sum_{r=1}^{\alpha_{m+1}} \proj{e_r\,e_r^*}
\label{ros}
\end{equation}
where the components of $\ket{e_r} \in {\cal C}^m$ are
\begin{equation}
\braket{i}{e_r} \equiv \frac{\sqrt{a_i}}{(1+R)^{\frac{1}{4}}} \exp{(\frac{2\pi J}{\alpha_{m+1}} \alpha_ir}), \,\,\,\,\,\,\, (J \equiv \sqrt{-1}),
\end{equation}
$\braket{i}{e_r^*}$ is just the complex conjugate of $\braket{i}{e_r}$, and the coefficients $\alpha_j$ are defined by:
\begin{eqnarray}
\alpha_j &\equiv& 2\alpha_{j-1} + 1, \\
\alpha_1 &\equiv& 0.
\end{eqnarray}
 To see that $\rmes = \rho_s$, and that therefore  $\rmes$ is separable, consider the matrix element of $\sum_{r=1}^{\alpha_{m+1}}\proj{e_r\,e_r^*}$:

\begin{eqnarray}
\sum_{r=1}^{\alpha_{m+1}} \braket{ij}{e_r\,e_r^*}\!\braket{e_r\,e_r^*}{kl} =  \nonumber \\
\frac{\sqrt{a_ia_ja_ka_l}}{1+R} \sum_{r=1}^{\alpha_{m+1}} \exp{\{\frac{2\pi J r}{\alpha_{m+1}}  (\alpha_i+\alpha_k-\alpha_j-\alpha_l)\}}
\label{ijcomp}
\end{eqnarray}
 Now, since $0 \leq i, j, k, l < m+1$ (and recalling that $\alpha_{m+1} > 2\alpha_m$), the quantity $\mid\!\! \alpha_i + \alpha_k - \alpha_j - \alpha_l\!\!\mid$ is always smaller than $\alpha_{m+1}$. Taking this into account, and also the fact that
\begin{equation}
\alpha_i + \alpha_k - \alpha_j - \alpha_l = 0 \Longleftrightarrow \left\{ \begin{array}{c}
   \left. \begin{array}{ll} i = j \\ k = l \end{array}      \right\}\\
   and/or \\
   \left. \begin{array}{ll} i = l \\ j = k  \end{array}     \right\}
 \end{array}\right. 
\end{equation}
we are left with the only non-vanishing elements
\begin{equation}
\bra{ii} \ros \ket{jj} = \bra{ij}\ros\ket{ij} = \frac{a_ia_j}{1+R}.
\end{equation}
This proves $\ros^+ = \ros$ and thus that $\ros^+$ is separable.
\vspace{2mm}

 Let us now see that there is no separable state $\ros$ such that 

\begin{equation}
\frac{1}{1+t}(\proj{\Psi} + t\,\ros)
\end{equation}
is separable with $t < R$. 
Recall that a necessary condition for $\rho$ to be separable is that its partial transposition $\rho^{T_i}$ (in the Hilbert space ${\cal H}^i$ of party $i$, $i=A,B$ in this case) be non-negative \cite{Peres},  that is:

\begin{equation}
\rho \in {\cal S}\,\,\,\, \Longrightarrow \,\,\,\,\,\rho^{T_i} \geq 0 \,\,\,\,    \forall i.\label{condition}
\end{equation}
Then $\ros$ and $t$ must necessarily satisfy
\begin{equation}
\bra{\Phi}\,\,\,\frac{1}{1+t}(\,\proj{\Psi} + t\,\ros\,)^{T_B}\,\,\, \ket{\Phi} \geq 0.
\label{condicio}
\end{equation}
for any $\ket{\Phi} \in {\cal C}^m \!\otimes{\cal C}^m$. Define a set of Bell states:
\begin{eqnarray}
\ket{\Phi_{ij}^+} &\equiv& \frac{1}{\sqrt{2}}(\ket{ij}\,+\,\ket{ji});\,\,\, P_{ij}^+ \equiv \proj{\Phi_{ij}^+},\\
\ket{\Phi_{ij}} &\equiv& \frac{1}{\sqrt{2}}(\ket{ij}\,-\,\ket{ji});\,\,\, P_{ij} \equiv \proj{\Phi_{ij}}.
\end{eqnarray}
Then the spectral decomposition of $\proj{\Psi}^{T_B}$ can be expressed in terms of $\Phi_{ij}^+$, $\Phi_{ij}$ and $\ket{ii}$:
\begin{equation}
\proj{\Psi}^{T_B} = \sum_{i=1}^m a_i^2 \proj{ii}\,+\,\sum_{i=1}^m\sum_{j>i}^m a_ia_j (P_{ij}^+ - P_{ij}).
\end{equation}
Now, from Eq.(\ref{condicio}) for $\ket{\Phi} = \ket{\Phi_{ij}}$,
\begin{equation}
\bra{\Phi_{ij}}\,\,\,\frac{1}{1+t}(\,\proj{\Psi}^{T_B} + t\,\rost\,)\,\,\, \ket{\Phi_{ij}} \geq 0, 
\end{equation}
which immediately leads to
\begin{equation}
Tr(\,P_{ij}\,\rost \,) \geq \frac{a_ia_j}{t}.
\end{equation}
Then, taking into account that $\sum_{i,j>i}a_ia_j = \frac{1}{2}(\sum_ia_i^2 - 1)$, we get
\begin{equation}
\sum_{i,j>i} \, Tr(\,P_{ij}\,\rost \,) \geq \frac{R}{2t}.
\label{trace}
\end{equation}
It will next be proved that $\sum_{i,j>i} \, Tr(\,P_{ij}\,\rost \,) \leq \frac{1}{2}$, which implies, together with Eq.(\ref{trace}), that $t \geq R$. Thus, the proposed local pseudomixture is an optimal one, and $R(\Psi) = R$. 

 Define the projector $M \equiv \sum_{i,j>i} \,P_{ij}$ and consider a symmetric unitary product transformation $U_\alpha \!\otimes\! U_\alpha$.

\vspace{2mm} 
{\bf Theorem B.1}
\vspace{2mm} 

 $[M,\,U_\alpha\!\otimes\! U_\alpha] = 0$.

\vspace{2mm} 
{\bf Proof:} For any $i,j$, $M \ket{ij} = P_{ij} \ket{ij} = \frac{1}{2}(\ket{ij} - \ket{ji})$ and, if $U_\alpha \ket{i} = \sum_{i'}b^i_{i'}\ket{i'}$,
\begin{eqnarray}
M U_\alpha\!\otimes\! U_\alpha \ket{ij} = \sum_{i',j'} b^i_{i'}b^j_{j'}M\ket{i'j'} = \sum_{i',j'} b^i_{i'}b^j_{j'}\frac{1}{2}(\ket{i'j'}-\ket{j'i'}) \nonumber \\
 = U_\alpha\!\otimes\! U_\alpha \frac{1}{2}(\ket{ij}-\ket{ji}) = U_\alpha\!\otimes\! U_\alpha M\ket{ij}.  
\end{eqnarray}
 This proves the theorem, since $\{\ket{ij}\}_{i,j=1\cdots m}$ is a basis of the whole Hilbert space.$\,\,\,\Box$ 

\vspace{2mm} 
{\bf Theorem B.2} (Necessary condition for separability)
\vspace{2mm} 

\begin{equation}
\rho \in {\cal S}\,\,\,\, \Longrightarrow \,\,\,\, Tr[\,\rho M] \leq \frac{1}{2},
\label{newcond}
\end{equation}

{\bf Proof:} Recall that if $\rho$ is separable, then it can be expressed as a convex combination of (not necessarily orthogonal) projectors onto product vectors $\ket{f_kg_k}$, that is, $\rho = \sum_k p_k\proj{f_kg_k}$.  Consider the following quantity:
\begin{equation}
M_{fg} \equiv \bra{fg}\,M\,\ket{fg}.
\end{equation}
 It will be proved that $M_{fg} \leq \frac{1}{2}$ for any product vector $\ket{fg}$, and that therefore, 
\begin{equation}
Tr[\rho\, M] = \sum_k p_k M_{f_kg_k} \leq \frac{1}{2}.
\label{rtrace}
\end{equation}
 Indeed, by noticing that theorem B.1 implies that $M_{U_{\alpha}f\, U_{\alpha}g} = M_{fg}$, since
\begin{equation}
Tr[\,\, U_{\alpha}\!\otimes\! U_{\alpha}\,\,\proj{fg} \,\,U_{\alpha}^{-1}\!\!\otimes\! U_{\alpha}^{-1}\,\, M\,\,] = Tr[\proj{fg}M\,],
\end{equation}
instead of $M_{fg}$ we can compute $M_{1\tilde{g}}$, where
\begin{equation}
\ket{1\tilde{g}} \equiv \left(  \begin{array}{l} 1 \\ \left.\!\!\!\begin{array}{c} 0 \\... \\0 \end{array}\! \right\}\! \mbox{\footnotesize m-1}\!\!\!\! \end{array}  \right) \otimes  \left(  \begin{array}{l} \tilde{g}_1 \\ \tilde{g}_2 \\ ...\\ \tilde{g}_m\end{array}  \right)
\end{equation}
for some $\ket{\tilde{g}}=U_{\alpha}\ket{g}$, where $U_{\alpha}$ is such that $\ket{1}=U_{\alpha}\ket{f}$. Then,
\begin{equation}
\bra{1\tilde{g}}M \ket{1\tilde{g}} = \sum_{i,j>i}\bra{1\tilde{g}}P_{ij}\ket{1\tilde{g}} = \sum_{j=2}^m \bra{1\tilde{g}}P_{1j}\ket{1\tilde{g}} = \frac{1}{2} \sum_{j>=2}^m |\braket{j}{\tilde{g}}|^2 \leq \frac{1}{2}|\braket{\tilde{g}}{\tilde{g}}|^2 = \frac{1}{2}.\,\,\,\Box
\end{equation}

\vspace{2mm} 

 {\bf B.2} Now the result
\begin{equation}
R(\,\Psi(\{a_i\})\, || \frac{1}{n}I ) = na_1a_2, 
\end{equation}
where $n \equiv n_1n_2$ is the dimension of the Hilbert space ${\cal H} = {\cal C}^{n_1} \otimes {\cal C}^{n_2}$ of the binary system, follows straightforwardly from the previous considerations. Indeed, with $m \equiv \min(n_1,\, n_2)$, $R_r \equiv n_1n_2a_1a_2$ and $R$ given by Eq.(\ref{purs}) ($R_r \geq R$ by construction), the separable state $\frac{1}{n}I$ can be written as a convex combination of $\rmenys$ from Eq.(\ref{rmenys}) and another manifestly separable state $\tilde{\ros}$:
\begin{eqnarray}
\frac{1}{n}I = \frac{1}{n}\sum_{i=1}^{n_1}\sum_{j=1}^{n_2} \proj{ij} \nonumber \\
 =\frac{1}{R_r}(\sum_{i=1}^{m}\sum_{j=1}^{m} a_ia_j\proj{ij} + \sum_{i=1}^{n_1}\sum_{j=1}^{n_2} c_{ij} \proj{ij})  \nonumber \\
= \frac{1}{R_r}( R\rmenys + (R_r - R)\tilde{\ros}),
\end{eqnarray}
where 
\begin{equation}
c_{ij} \equiv  \left\{ \begin{array}{ll} 
a_1a_2- a_ia_j \,\,(\geq 0) \,\,\,\,\,  \mbox{ if }\, i,j \leq m\\
a_1a_2 \hspace{26mm}\,\,\mbox{otherwise},
\end{array}\right.
\end{equation}
and 
\begin{equation}
\tilde{\ros} \equiv \frac{1}{R_r - R}  \sum_{i=1}^{n_1}\sum_{j=1}^{n_2} c_{ij} \proj{ij} \in {\cal S}.
\label{rostilde}
\end{equation}
 Then $\frac{1}{1+R_r}(\proj{\Psi}+R_r\frac{1}{n}I) = \frac{1}{1+ R_r}((1+R)\rmes + (R_r-R)\tilde{\ros})$, where $\rmes$ was defined in Eq.(\ref{rmes}), is manifestly separable, whereas one could check that for any $\epsilon > 0$
\begin{equation}
\bra{\Phi_{12}} \,\,\proj{\Psi}^{T_B} + (R_r-\epsilon)\frac{1}{n}I^{T_B}\,\,\ket{\Phi_{12}} = -\frac{\epsilon}{n} < 0, 
\end{equation}
so that, recalling the necessary condition for separability discussed in Eq.(\ref{condition}), $R_r$ is the minimum amount of $\frac{1}{n}I$ that mixed with $\proj{\Psi}$ makes it separable, that is $R(\Psi||\frac{1}{n}I) = R_r$.

\vspace{2mm} 

 {\bf B.3} Let us finally consider, as an example, a pure state of the smallest composite system: a system of two qubits. In this case the Hilbert space is ${\cal C}^2 \otimes {\cal C}^2$, and the ordered Schmidt decomposition allows us to express with an adequate choice of basis any pure state $\Psi$ as
\begin{equation}
\ket{\Psi} = a_1\ket{1}\!\otimes\!\ket{1} + a_2\ket{2}\!\otimes\!\ket{2} = \left( \begin{array}{c}
a_1 \\
0 \\
0 \\
a_2
\end{array} \right).
\end{equation}
Then, using the definitions given in B.1, $R = 2a_1a_2$,
\begin{equation}
\rmenys \equiv \frac{1}{2} (\proj{12}+\proj{21}) =  \frac{1}{2} \left( \begin{array}{llll}
0 & 0 & 0 & 0 \\
0 & 1 & 0 & 0 \\
0 & 0 & 1 & 0 \\
0 & 0 & 0 & 0 
\end{array} \right),
\end{equation}
and
\begin{equation}
\rmes = \frac{1}{1+R} \left( \begin{array}{cccc}
a_1^2 & 0 & 0 & a_1a_2  \\
0 & a_1a_2 & 0 & 0 \\
0 & 0 & a_1a_2 & 0 \\
a_1a_2 & 0 & 0 & a_2^2 
\end{array} \right). 
\label{exemplermes}
\end{equation}
To check that $\rmes = \ros$ as given by Eq. (\ref{ros}) let us specify it for our example:
\begin{equation}
\rho_s =  \frac{1}{3} \sum_{k=1}^{3} \proj{e_r\,e_r^*},
\end{equation}
where
\begin{equation}
 \ket{e_r\,e_r^*} \equiv \left( \begin{array}{l} a_1^\frac{1}{2} \\ a_2^\frac{1}{2}\exp{\{\frac{2\pi J}{3}r\}} \end{array} \right) \otimes \left( \begin{array}{l} a_1^\frac{1}{2} \\ a_2^\frac{1}{2}\exp{\{\frac{\!-\!2\pi J}{3}r\}} \end{array} \right).
\end{equation}
 More explicitly,
\begin{equation}
\rho_s =  \frac{1}{3(1+R)} \sum_{r=1}^{3} N_r,
\end{equation}
with $N_r$ given by
\begin{equation}
  \left( \begin{array}{cccc}
a_1^2 & a_1^\frac{3}{2}a_2^\frac{1}{2}\exp{\{\frac{2\pi J}{3}r\} } & a_1^\frac{3}{2}a_2^\frac{1}{2} \exp{\{\frac{\!-\!2\pi J}{3}r\}} & a_1a_2  \\
a_1^\frac{3}{2}a_2^\frac{1}{2} \exp{\{\frac{\!-\!2\pi J}{3}r\}} & a_1a_2 & a_1a_2 \exp{\{\frac{\!-\!2\pi J}{3}2r\}} & a_1^\frac{1}{2}a_2^\frac{3}{2} \exp{\{\frac{\!-\!2\pi J}{3}r\}} \\
a_1^\frac{3}{2}a_2^\frac{1}{2}\exp{\{\frac{2\pi J}{3}r\} } & a_1a_2 \exp{\{\frac{2\pi J}{3}2r\}} & a_1a_2 & a_1^\frac{1}{2}a_2^\frac{3}{2}\exp{\{\frac{2\pi J}{3}r\} }  \\
a_1a_2 & a_1^\frac{1}{2}a_2^\frac{3}{2}\exp{\{\frac{2\pi J}{3}r\} } & a_1^\frac{1}{2}a_2^\frac{3}{2} \exp{\{\frac{\!-\!2\pi J}{3}r\}} & a_2^2 
\end{array} \right).
\end{equation}
 The sum over $r$ now reproduces Eq.(\ref{exemplermes}) immediately so that $\rho_s = \rmes$. Some of the expressions used in proving that the local pseudomixture $\{\rmes, \rmenys, R\}$ is optimal read for our example: 
\begin{equation}
\proj{\Psi}^{T_B} = a_1^2 \proj{11}\,+a_2^2 \proj{22}\,+\,a_1a_2 (P_{12}^+ - P_{12}) =  \left( \begin{array}{cccc}
a_1^2 & 0 & 0 & 0  \\
0 & 0 & a_1a_2 & 0 \\
0 & a_1a_2 & 0 & 0 \\
0 & 0 & 0 & a_2^2 
\end{array} \right),
\end{equation}
and
\begin{eqnarray}
\ket{\Phi_{12}^+} = \frac{1}{\sqrt{2}} \left( \begin{array}{c}
0 \\ 1 \\ 1 \\ 0 \end{array} \right),
\,\,\,\ket{\Phi_{12}} = \frac{1}{\sqrt{2}} \left( \begin{array}{l}
0 \\ 1 \\\!\!\!-\!1 \\ 0 \end{array} \! \right),
\end{eqnarray}
and Eq.(\ref{trace}) is
\begin{equation}
Tr(\,P_{12}\,\rost \,) \geq \frac{a_1a_2}{t},
\end{equation}
which, taking into account that $\bra{fg}\,P_{12}\,\ket{fg} \leq \frac{1}{2}$ for any product vector $\ket{fg}$ (theorem B.2), and consequently $Tr(\,P_{12}\,\rho\,) \leq \frac{1}{2}$ for any separable $\rho$, implies that $R(\Psi)=R=(a_1+a_2)^2-1$.

Now $R_r=4a_1a_2$ and the maximally random state in ${\cal C}^2 \otimes {\cal C}^2$ can be decomposed, following B.2, as a mixture of two separable states as follows:
\begin{equation}
\frac{1}{4}I = \frac{1}{R_r} \{\left( \begin{array}{cccc}
0 & 0 & 0 & 0  \\
0 & a_1a_2 & 0 & 0 \\
0 & 0 & a_1a_2 & 0 \\
0 & 0 & 0 & 0 
\end{array} \right) +
 \left( \begin{array}{cccc}
a_1a_2 & 0 & 0 & 0  \\
0 & 0 & 0 & 0 \\
0 & 0 & 0 & 0 \\
0 & 0 & 0 & a_1a_2 
\end{array} \right) \},
\end{equation}
so that
\begin{equation}
\tilde{\ros} = \frac{1}{2} \left( \begin{array}{cccc}
1 & 0 & 0 & 0  \\
0 & 0 & 0 & 0 \\
0 & 0 & 0 & 0 \\
0 & 0 & 0 & 1 
\end{array} \right),
\end{equation}
in Eq.(\ref{rostilde}).

\vspace{4mm}

\renewcommand{\theequation}{C.\arabic{equation}}
\setcounter{equation}{0}

{\bf Appendix C: Mainly bounds for robustness and random robustness of mixed states of binary systems.}
\vspace{4mm}

\vspace{2mm}
{\bf C.1} $\,\,\,\,|\min(\{\frac{\lambda_j}{a_{j,1}^2}\},0)| \leq R(\rho)$
\vspace{2mm}

 {\bf Proof:} Assume that in the spectral decomposition of $\rho^{T_B}$, $\rho^{T_B} = \sum_{j=1}^{n} \lambda_j \proj{\Psi_j}$, at least one eigenvalue, say $\lambda_j$, is negative. Calling the non-local coefficients of the ordered Schmidt decomposition of the corresponding eigenvector, $\ket{\Psi_j}$, $\{a_{j,i}\}$, one finds that, for $\ros \in {\cal S}$, if $\bra{\Psi_j}\, (\rho + t\ros)^{T_B}\, \ket{\Psi_j} $ is to be non-negative (which is a necessary condition for $\frac{1}{1+t}(\rho + t\ros)$ to be separable), then $t \geq \frac{-\lambda_j}{\bra{\Psi_j}\ros^{T_B}\ket{\Psi_j}}$. We will next prove that $|\braket{\Psi_j}{p}| \leq a_{j,1}$
 for any product vector $\ket{p}$, and therefore $\bra{\Psi_j}\ros^{T_B}\ket{\Psi_j} \leq a_{j,1}^2$, which implies the lower bound for the robustness of $\rho$.  If, on the contrary, no $\lambda_j < 0$ exists, no significant bound is obtained.

\vspace{2mm}
{\bf Theorem C.1}
\vspace{2mm}

 If $\ket{\Psi} = \sum_{i=1}^m a_i \ket{i}\otimes\ket{i}$ is the ordered Schmidt decomposition of the normalised vector $\ket{\Psi} \in {\cal C}^{n_1} \!\otimes{\cal C}^{n_2}$ (i.e. $m = \min(n_1,n_2)$, $a_i \geq a_{i+1} \geq 0$ and $\{\ket{i}\}_{i=1,...m}$ are othonormal vectors) and $\ket{p} \equiv \ket{p_1}\otimes\ket{p_2} \in {\cal C}^{n_1} \!\otimes{\cal C}^{n_2}$ is any normalised product vector, then $|\braket{\Psi}{p}|\leq a_1$.

{\bf Proof:} For $p_{1,i} \equiv \braket{i}{p_1}$ and $p_{2,i} \equiv \braket{i}{p_2}$, one gets
\begin{eqnarray}
|\braket{\Psi}{p}| = |\sum_{i=1}^m a_ip_{1,i}p_{2,i}| \leq \sum_{i=1}^m a_i|p_{1,i}p_{2,i}|
 \leq a_1 \sum_{i=1}^m |p_{1,i}p_{2,i}| \\
\leq a_1 \sqrt{\sum_{i=1}^m |p_{1,i}|^2} \sqrt{\sum_{i=1}^m |p_{2,i}|^2} \leq a_1 \sqrt{\braket{p}{p}} = a_1. \,\,\,\Box
\end{eqnarray}

\vspace{2mm}
{\bf C.2} $\,\,\,R(\rho) \leq \tilde{m}-1$, where $\tilde{m} = \min(\mbox{rank}[\rho^A],\mbox{rank}[\rho^B])$ 
\vspace{2mm}

{\bf Proof:} For $\tilde{\cal H}\subseteq{\cal H}$ the product subspace spanned by the eigenvectors of $\rho^A\otimes\rho^B$ with non-vanishing eigenvalue, any rank one projector in a convex combination of $\rho$ happens to project into $\tilde{\cal H}$, that is, if $\rho = \sum p_k \proj{\Psi_k}$, then $\ket{\Psi_k} \in \tilde{\cal H}$. But $R(\Psi_k) = (\sum_{i=1}^{\tilde{m}} a_i)^2-1 \leq (\sum_{i=1}^{\tilde{m}} \frac{1}{\sqrt{\tilde m}})^2-1 = \tilde{m} - 1$. Then, since $R(\rho)$ is a convex function, $R(\rho) \leq \sum p_k R(\Psi_k) \leq \tilde{m} -1$.

\vspace{2mm}
{\bf C.3} $R(\rho) \leq R(\tilde{\rho}||\frac{1}{\tilde n}\tilde{I})$
\vspace{2mm}
 follows from the fact that $\frac{1}{\tilde n}\tilde{I}$ is a separable state, and $R(\tilde{\rho})=R(\rho)$ is the minimum of the relative robustness $R(\tilde{\rho}||\rho_s)$.

\vspace{2mm}
{\bf C.4} $R(\rho) \leq 2|\lambda|$ \hspace{4mm} (${\cal C}^{2} \!\otimes{\cal C}^{2}$)
\vspace{2mm}

{\bf Proof:} The partially transposed $\rho^{T_B}$ of any inseparable density matrix $\rho$ in ${\cal C}^{2} \!\otimes{\cal C}^{2}$ has always a negative eigenvalue $\lambda$ \cite{STV}, for a certain eigenvector $\ket{n} = \cos{\theta} \ket{11} +  \sin{\theta} \ket{22}$. (Here we choose the local basis $\{\ket{ij}\}_{i,j=1,2}$ to be that defined by the Schmidt decomposition of $\ket{n}$). For $\ros \equiv \cos^2 \theta \proj{11} + \sin^2 \theta \proj{22}$ it can be checked that $-\proj{n} + 2\ros \geq 0$, which implies that 
\begin{equation}
\frac{1}{1+2|\lambda|} (\rho + 2|\lambda|\ros)
\end{equation}
is a separable state.

\vspace{2mm}
{\bf C.5} $R(\rho) \leq C(\rho)$ \hspace{4mm} (${\cal C}^{2} \!\otimes{\cal C}^{2}$)
\vspace{2mm}

{\bf Proof:} The robustness $R(\Psi)$ and the concurrence $C(\Psi)$ are equal for any pure state of ${\cal C}^{2} \!\otimes{\cal C}^{2}$, and in \cite{Wootters} it was proved that one can always find a realization $\{\Psi_k, p_k\}$ of four pure states for $\rho$ such that $C(\Psi_k) = C(\rho) \,\,\, \forall k$. Then, using the convexity of $R(\rho)$, we find that for this realization
\begin{equation}
R(\rho) \leq \sum_{k=1}^4 p_k R(\Psi_k) = \sum_{k=1}^4 p_k C(\Psi_k) = C(\rho).
\end{equation}

\vspace{2mm}
{\bf C.6} $R(\rho(p,q_1,q_2,\theta)) = (1-p)\sin{2\theta} - pq_2\,\,\,$ if $\rho^{T_B} \not{\!\!\geq} 0$\hspace{4mm} (${\cal C}^{2} \!\otimes{\cal C}^{2}$)
\vspace{2mm}

{\bf Proof:} $\bra{\Phi}\rho^{T_B}\ket{\Phi} = \frac{1}{2}(pq_2 - (1-p)\sin{2\theta})$ for $\ket{\Phi} \equiv \frac{1}{\sqrt{2}} (\ket{12}-\ket{21})$. Then a necessary condition for 
\begin{equation}
\frac{1}{1+t}(\rho + t\ros)
\label{tipic}
\end{equation}
to be separable for a separable $\ros$ is that $\bra{\Phi}(\rho+t\ros)^{T_B}\ket{\Phi} \geq 0$, that is $-\bra{\Phi}\rho^{T_B}\ket{\Phi} \leq t \bra{\Phi}\ros^{T_B}\ket{\Phi}$. But in Appendix B it was proved that $\bra{\Phi}\ros^{T_B}\ket{\Phi} \leq \frac{1}{2}$, so that $t \geq (1-p)\sin{2\theta} - pq_2$. Moreover one can check that $\ros \equiv \frac{1}{2}{\proj{12} + \proj{21}}$ with weight $t = (1-p)\sin{2\theta} - pq_2$ makes the density matrix in Eq.(\ref{tipic}) separable. 

\vspace{2mm}
{\bf C.7} $R(\rho||\frac{1}{n}I) \geq n|\min (\{\lambda_k\}, 0)|$ 
\vspace{2mm}

{\bf Proof:} For any $\rho$ consider the spectral decomposition of $\rho^{T_B}$
\begin{equation}
\rho^{T_B} = \sum_{k=1}^n \lambda_k \proj{\Psi_k},
\end{equation}
where we take $\lambda_k \leq \lambda_{k+1}$ (and we take into account also eigenvectors with vanishing eigenvalue), and suppose $\lambda_1 < 0$. Then $(\rho + n|\lambda_1|\frac{1}{n}I)^{T_B} = \sum_{k=1}^n (|\lambda_1|+\lambda_k) \proj{\Psi_k}$ is manifestly non-negative definite (which is a necessary condition for the separability of $\frac{1}{1+n|\lambda_1|}(\rho + n|\lambda_1|\frac{1}{n}I)$), whereas for any $\epsilon > 0$, 
\begin{equation}
\bra{\Psi_1}(\rho + (n|\lambda_1|-\epsilon)\frac{1}{n}I)^{T_B}\ket{\Psi_1} = -\frac{\epsilon}{n} < 0.
\end{equation}
 If $\lambda_1 \geq 0$ no significant bound is obtained.
\vspace{2mm}

{\bf C.8} For $N=2$ the bound $R(\rho||\frac{1}{n}I) \leq \frac{n}{2}$ is a consequence of the fact that for any pure state of a binary system $R(\Psi||\frac{1}{n}I) = na_1a_2 \leq \frac{n}{2}$, and of the convexity of $R(\rho||\frac{1}{n}I)$ as a function of $\rho$, that the reader can easily prove. Its generalization to $N$-party systems,
\begin{equation}
R(\rho||\frac{1}{n}I) \,\, \leq \,\, \left(1+\frac{n}{2}\right)^{N\!-\!1}-1,
\label{cota}
\end{equation}
can be derived from the previous result and we will explain it only for $N=3$, the $N>3$ case following straighforwardly. Consider a pure state $\Psi^{ABC}$ shared by Alice, Bob and Claire. If we first think of Bob and Claire as a single party, then we have seen that the state 
\begin{equation}
\frac{1}{1+\frac{n}{2}}(\proj{\Psi^{ABC}} + \frac{n}{2}\frac{1}{n}I) 
\label{pseudosep}
\end{equation}
is separable if considered as belonging to a binary system, that of Alice as one party and Bob and Claire as the other, and therefore can be expressed as a convex combination $\sum_{k}p_k\proj{\psi_k^A}\otimes\proj{\phi^{BC}_k}$ of pure states that are product in ${\cal H}^A\otimes {\cal H}^{BC}$. Now mixing any of these pure states with an amount $\frac{n}{2}$ of $\frac{1}{n}I$ we obtain a proper separable state:
\begin{eqnarray}
\frac{1}{1+\frac{n}{2}}\left(\proj{\psi^A}\otimes\proj{\phi^{BC}} + \frac{n}{2}\frac{1}{n}I\right) = \\ 
\frac{1}{1+\frac{n}{2}}\left(\proj{\psi^A}\otimes(\proj{\phi^{BC}} +  \frac{n}{2}\frac{1}{n}I^{BC}) 
+ \frac{n}{2}\frac{1}{n}(I^A-\proj{\psi^A})\otimes I^{BC}\right),
\label{general}
\end{eqnarray}
where $I^i$ is the identity matrix in ${\cal H}^i$. Indeed, $\frac{1}{1+\frac{n_Bn_C}{2}}(\proj{\phi^{BC}} +  \frac{n}{2}\frac{1}{n}I^{BC})$ is a separable state in ${\cal H}^B\otimes {\cal H}^{C}$, whereas $\frac{1}{n_A-1}(I^A-\proj{\psi^A})$ is a mixed state in ${\cal H}^{A}$, so that the RHS of Eq.(\ref{general}) is a convex combination of two manifestly separable states. Then, by adding an amount  $\frac{n}{2}$ of the separable $\frac{1}{n}I$ to the state in Eq.(\ref{pseudosep}) we make it separable, and therefore mixing the initial pure state $\Psi^{ABC}$ with an amount $\frac{n}{2} + (1+\frac{n}{2})\frac{n}{2}=(1\!+\!\frac{n}{2})^2\!-\!1$ of $\frac{1}{n}I$ is sufficient to wash out its quantum correlations.

\end{document}